\documentclass[preprint,a4paper,prd,preprintnumbers,amssymb,superscriptaddress,nofootinbib,numerical,tightenlines]{revtex4-1} 
\usepackage[english]{babel}
\usepackage{pdfsync}	
\usepackage[text={17.0cm,23.0cm}]{geometry}
\usepackage{graphicx,epsfig,graphics,verbatim}
\usepackage{mathrsfs,amsmath,amssymb,amsbsy}
\usepackage{slashed}
\usepackage{psfrag}
\usepackage{color}
\usepackage[normalem]{ulem}
\usepackage[utf8]{inputenc}
\usepackage{hyperref}
\usepackage{url}
\usepackage{subcaption}
\usepackage{blindtext}

\usepackage[font=scriptsize,justification=raggedright,skip=15pt]{caption}
\DeclareCaptionLabelFormat{capital-short}{FIG. #2}
\def\lsim{\mathrel{\rlap{\lower4pt\hbox{\hskip1pt$\sim$}}
    \raise1pt\hbox{$<$}}}                
\def\gsim{\mathrel{\rlap{\lower4pt\hbox{\hskip1pt$\sim$}}
    \raise1pt\hbox{$>$}}}                
\newcommand{\be}{\begin{equation}}
\newcommand{\ee}{\end{equation}}
\newcommand{\epsl}{\slashed{\epsilon}^*}
\newcommand{\epst}{{\epsilon}^*}
\newcommand{\ksl}{\slashed{k}}

\newcommand{\usp}{\bar{u}(p_1)}
\newcommand{\vsp}{{v}(p_2)}

\newcommand{\avsv}{\langle \sigma v \rangle}
\newcommand{\MP}{M_{\psi}}

\newcommand{\Mfsr}{\mathcal{M}_{\rm FSR}}
\newcommand{\Mvib}{\mathcal{M}_{\rm VIB}}
\newcommand{\Mib}{\mathcal{M}_{\rm IB}}
\newcommand{\itemmul}[2]{#1\mkern1mu{\cdot}\mkern1mu#2}
\newcommand{\Ieik}{\mathcal{I}_{\rm eik.}}

\newcommand{\qqg}{q \overline{q} g}
\newcommand{\bo}{\beta_0}
\definecolor{dgreen}{rgb}{0.,0.6,0.}

\begin{document}
\preprint{ULB-TH/18-08}
\preprint{BONN-TH-2018-08}
  \title{On Radiative Corrections to Vector-like Portal Dark Matter}

\author{Stefano Colucci}
\email{colucci@th.physik.uni--bonn.de}
\affiliation{Physikalisches Institut der Universit\"at Bonn, Bethe Center
for Theoretical Physics, \\ Nu{\ss}allee 12, 53115 Bonn, Germany}

\author{Federica Giacchino}
\email{federica.giacchino@ulb.ac.be}
\affiliation{Service de Physique Th\'eorique, CP225, Universit\'e Libre de Bruxelles, Bld du Triomphe, 1050 Brussels, Belgium}

\author{Michel H.G.~Tytgat}
\email{mtytgat@ulb.ac.be}
\affiliation{Service de Physique Th\'eorique, CP225, Universit\'e Libre de Bruxelles, Bld du Triomphe, 1050 Brussels, Belgium}

\author{J\'{e}r\^{o}me Vandecasteele}
\email{jerome.vandecasteele@ulb.ac.be}
\affiliation{Service de Physique Th\'eorique, CP225, Universit\'e Libre de Bruxelles, Bld du Triomphe, 1050 Brussels, Belgium}

\date{\today}

\begin{abstract}
A massive real scalar dark matter particle $S$ can couple to Standard Model leptons or quarks through a vector-like fermionic mediator $\psi$, a scenario known as the Vector-like portal. Due to helicity suppression of the annihilation cross section into a pair of SM fermions, it has been shown in previous works that radiative corrections, either at one-loop or through radiation of gauge bosons, may play a significant role both in determining the relic abundance and for indirect detection. All previous works considered the limit of massless final state quarks or leptons. 
In this work, we focus on a technical issue, which is to reliably determine the annihilation cross sections taking into account finite fermion masses. Following previous works in the framework of simplified supersymmetric dark matter scenarios, and building on an analogy with Higgs decay into fermions, we address the issue of infrared and collinear divergences that plagues the cross section by adopting an effective operator description, which captures most of the relevant physics and give explicit expressions for the annihilation cross sections. We then develop several approximations for the differential and total cross sections, which simplify greatly their expressions, and which can then be used in various phenomenological studies of similar models. Finally, we describe our method  to compute the final gamma-ray spectrum, including hadronisation of the heavy fermions, and provide some illustrative spectra for specific dark matter candidates. 
\end{abstract}

\maketitle


\section{Introduction}
Dark matter (DM) amounts to about 27\% of the energy budget of our universe and, yet, little is known about its precise nature. A much studied possibility is that dark matter is made of a weakly interacting massive particle (WIMP). In this scenario the observed relic abundance $\Omega_\text{DM} h^2 = 0.1186 \pm 0.0020$ ~\cite{Olive:2016xmw} more or less naturally results from chemical {freeze-out} of a massive particle, provided its annihilation cross section is $ \avsv \simeq 3 \cdot 10^{-26} {\rm cm}^{3}\cdot {\rm s}^{-1} $. Many such candidates have been proposed in the literature. 

In the present note, we study further a real scalar particle coupled to Standard Model (SM) fermions through a  vector-like fermion. This scenario has been dubbed the Vector-like Portal (VLP) in~\cite{Giacchino:2013bta} following \cite{Perez:2013nra}. The simplest realization of the VLP is given by
\begin{equation} \label{eq:LagrangianDM}
\mathcal{L}_{\rm DM}= - \left(y_f \, S\, \bar{\psi}\, f_R\, + \,{\rm h.c.}\right)  - \frac{1}{2} M_S^2 S^2 - M_\psi \bar \psi \psi \;,
\end{equation}
where $S$ is a singlet real scalar (the DM candidate), $f_R$ is an $SU(2)_L$ singlet SM fermion (lepton or quark) and $\psi$ a vector-like massive fermion. Clearly other combinations of SM multiplets could be considered, but this is not our purpose here. Stability of DM is ensured by imposing a discrete $Z_2$ symmetry, under which both $S$ and $\psi$ are taken to be odd and SM particles even. For simplicity and to avoid addressing flavour physics aspects, it is assumed that $S$ couples dominantly to a single SM flavour. Also, in the sequel we will assume that the possible quartic coupling of $S$ with the SM Higgs is small and may be neglected. As such, (\ref{eq:LagrangianDM}) falls in the category of so-called simplified DM models with a t-channel mediator, see {\em e.g.} \cite{Abdallah:2015ter,Abercrombie:2015wmb,Bauer:2016gys,DeSimone:2016fbz}. 

This model may be considered as the scalar version of a bino-like Majorana DM candidate, with which it shares some basic properties, the first being that their s-wave annihilation is helicity suppressed. They differ in the fact that annihilation of a bino-like candidate is p-wave in the chiral limit \cite{Goldberg:1983nd} while that of the real scalar $S$ is d-wave \cite{Toma:2013bka,Giacchino:2013bta}. However, in both cases, the helicity suppression is lifted by radiative corrections~\cite{Bergstrom:1989jr,Flores:1989ru}. As discussed in several works, this has interesting phenomenological implications. In the case of coupling to leptons, radiative processes, either in the form of internal bremsstrahlung or annihilation at one-loop into, say, two gamma-rays, may lead to striking spectral features. Such spectral features are of interest for indirect searches for WIMPs (see, e.g.\cite{Bringmann:2007nk,Bringmann:2012vr,Bell:2011eu,Garny:2013ama,Garny:2015wea} for the Majorana case and specifically \cite{Toma:2013bka,Giacchino:2013bta,Giacchino:2014moa,Ibarra:2014qma} for the scalar case). In the case of coupling to (light) quarks, radiative processes involving gluons on top of gammas may be relevant at the time of thermal freeze-out, thus impacting both the effective annihilation cross section and indirect signatures, see e.g. \cite{Giacchino:2015hvk,Bringmann:2015cpa}. 

 In previous works, annihilation of the particle $S$ through internal bremsstrahlung was only considered in the chiral limit, neglecting the mass of the final state fermions. This was motivated by simplicity, but also by physics. Indeed, it is for light leptons and quarks that radiative corrections are more important, by lifting the helicity suppression. Also, it is in this limit that they lead to most spectacular spectral signatures, with a sharp gamma-ray spectral feature around $E_\gamma \sim m_{\rm dm}$ when bremsstrahlung is dominated by emission from the intermediate particle\footnote{The separation into final state radiation (FSR) and VIB processes is not clear cut because of gauge invariance. However it becomes manifest in an expansion into effective operators, see section \ref{sec:radiative1}.}, a process called virtual internal bremsstrahlung (VIB) \cite{Bringmann:2007nk}. Instead, in this work, we want to consider the possibility that $S$ couples dominantly to heavy fermions. The prime application would be annihilation into a top-antitop pair, but we will also consider the indirect detection signatures from annihilation into $b\bar b$ and $\tau^+\tau^-$. 
 
The detailed phenomenological analysis of a top-philic candidate, including searches at the LHC and constraints from direct detection searches, is the object of a separate article \cite{Colucci:2018vxz}. In the present work, we specifically focus on more technical aspects of determining the total cross section, taking into account radiative corrections, as well as the spectra into gamma-rays relevant for indirect searches. Concretely, our goal is to keep track of the non-zero quark mass effects, the most important being that the s-wave part of the annihilation cross section into quark-antiquark is helicity suppressed. The issue we will have to face is that the total annihilation cross section is plagued by infrared (IR) divergences, associated to final state radiation (FSR) of soft gluons or gammas. According to the Kinoshita-Bloch-Nordsieck theorem~\cite{Kinoshita:1962ur,Bloch:1937pw} the full cross section is free of IR divergence. This involves properly taking into account radiative corrections at a given order in the gauge coupling. For the case at hand, this requires calculating the one-loop corrections to the annihilation cross section $SS \rightarrow f \bar f$. Although in principle straightforward, the calculations are involved. 
 
 In this note, we give a calculation of the annihilation cross section at next-to-leading order (NLO) following an effective approach advocated in \cite{Bringmann:2015cpa}, which they applied to the case of bino-like DM (see also \cite{Bringmann:2017sko} for more general cases ). The expression of the cross section is free of IR divergences and may be applied to the case of annihilation of $S$ in heavy quarks.  The main idea behind \cite{Bringmann:2015cpa} is to consider separately the emission of FSR and VIB gluons or gammas. The former is dominated by emission from final state fermions, which is the source for infrared divergences of the total cross section. 
 Following the Weizs\" acker-Williams approximation, the amplitude for emission of soft modes is obtained by multiplying the leading order (LO), tree-level amplitude by a universal factor (for fermions in the final state). For non-relativistic DM in an s-wave, this tree-level amplitude can be equivalently obtained from an effective contact interaction, which, in the scalar case, is given by the following 5-dimensional operator, where $m_q$ is the quark mass,
\begin{equation}
\label{eq:LOsoft} 
{\cal O}^{(5)}_{m} = \frac{m_q}{\Lambda^2} S^2 \bar q q\;.
\end{equation}
Consequently, the IR divergence can be tackled by taking into account the one-loop correction to the effective interaction of (\ref{eq:LOsoft}). The rest, that is the emission of hard modes, is IR safe, and can be obtained by considering the full, underlying theory. This effective approach simplifies very much the calculations, while capturing the underlying physics, {\em i.e.} with limited error compared to a full NLO calculation \cite{Bringmann:2015cpa}. The separation between soft and hard modes is implemented by a cut-off on the energy of the emitted gamma or gluon. This strategy has been used for calculating NLO QCD corrections to the decay of the Higgs \cite{Drees:1990dq,Braaten:1980yq}. We will follow closely the approach of Ref.\cite{Drees:1990dq}. Incidentally, for the soft part, the calculations are precisely the same as in the case of Higgs decay. They differ for the emission of hard modes, for which we will give complete expressions in the case of annihilating scalar dark matter. 

\bigskip

The manuscript is organized as follows. In section \ref{sec:radiative1}, we provide the calculation of the annihilation cross section of a real scalar DM particle into SM fermions through t-channel exchange of a vector-like fermion. We introduce the effective approach of \cite{Bringmann:2015cpa} and give explicit expressions from the decomposition of the cross section for soft emission, the associated one-loop corrections, and hard emission, the latter including virtual internal bremsstrahlung. In section \ref{sec:diff}, we study the differential cross sections (with gluon and gamma emission) and implications for indirect detection, in particular for the gamma-ray spectra. We  draw our conclusions in section \ref{sec:con}. Some lengthy expressions are relegated to the Appendices \ref{app:sv} and  \ref{app:limits}.

\section{Total annihilation cross section}
\label{sec:radiative1}

In this section, we first revisit the basics of the model, and the reason why internal bremsstrahlung may be relevant. We then go on with the main steps of the calculation of the cross section for massive final state SM fermions.

\subsection{Leading order annihilation cross section}
\begin{figure}[t]
\centering
\begin{subfigure}[b]{0.2\textwidth}
\includegraphics[width=\textwidth]{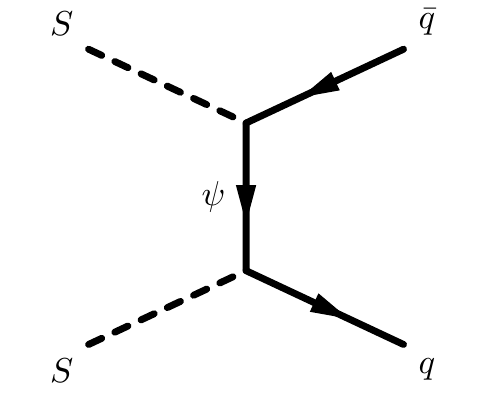}
\end{subfigure}
\quad\raisebox{1.2cm}{$\pmb{+}$}\quad
\begin{subfigure}[b]{0.2\textwidth}
\includegraphics[width=\textwidth]{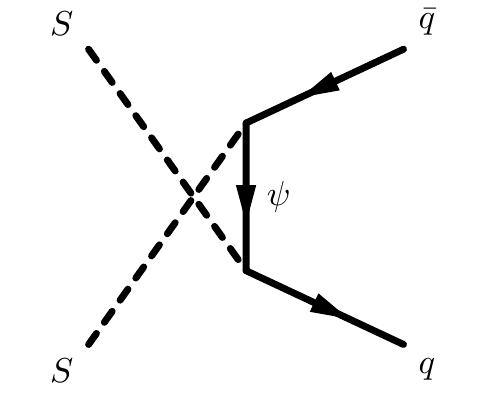}
\end{subfigure}
\quad\raisebox{1.2cm}{$\pmb{\longrightarrow}$}
\begin{subfigure}[b]{0.2\textwidth}
{\includegraphics[width=\textwidth]{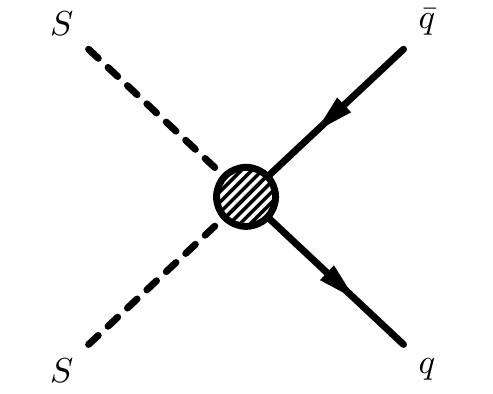}}
\end{subfigure}
\,\raisebox{1.2cm}{$\pmb{=} \frac{i}{2}\frac{y_f^2}{M_S^2(1+r-z)}$}
\caption{Amplitudes for the 2-body process $SS \to q\bar{q}$ and resulting effective interaction ($r = \MP^2/M_S^2$ and $z = m_q^2/M_S^2$).}
\label{fig:SStoqq}
\end{figure}

 We consider the amplitudes depicted by the Feynman diagrams in Fig.~\ref{fig:SStoqq}. 
 The annihilation cross section of non-relativistic DM particles is then
\be \label{eq:SStoqq}
\sigma v_{q \bar q}  = \frac{y_f^4 N_c }{4 \pi M_S^3} \frac{m_q^2\, (M_S^2 - m_q^2)^{3/2}}{(M_S^2+\MP^2- m_q^2)^2} +\mathcal{O}(m_q ^2 v^2, v^4)\;,
\ee
where $v$ is their relative velocity and $N_c$ the number of colours. 
The helicity suppression ($\sim m_q^2$) of the s-wave part of the cross section stems from the fact that, from Eq.~\eqref{eq:LagrangianDM}, the $S$ coupling to SM fermions is chiral while the quark-antiquark pair must have zero total helicity; matching the two requires a chirality flip.\footnote{{The power of $3/2$ of the phase-space factor $(1 - m_q^2/M_S^2)^{3/2}$ reveals that the final state quark-antiquark pair is in a p-wave near threshold. The reason is the same as for Higgs decay into a fermion-antifermion pair, see e.g. \cite{Drees:1990dq}. The scalar DM pair in an s-wave corresponds to a $J^{PC} = 0^{++}$ initial state. As the parity of the final state quark-antiquark pair is $P= - (-)^{l}$ (the minus factor is intrinsic parity), they must be in a p-wave. By the same token, they must have total spin $S=1$ to make a $J=0$ state. Since $C = (-)^{l+s}$ the final state is indeed $0^{++}$.  
A similar argument holds for s-wave annihilation of a pair of Majorana DM \cite{Bringmann:2015cpa}. The difference is that the initial state is instead in a $J^{PC} = 0^{-+}$, equivalent to a pseudo-scalar particle. Another, but related difference is that the cross section for scalar DM is d-wave suppressed in the chiral limit $m_q \rightarrow 0$, while it is p-wave in the Majorana case \cite{Toma:2013bka,Giacchino:2013bta}.}} Incidentally, the s-wave part of the cross section can be derived from the low-energy effective interaction in Eq.~\eqref{eq:LOsoft} with
\begin{equation}
\frac{m_q}{\Lambda^2} \rightarrow \frac{1}{2}\frac{y_f^2 m_q}{M_S^2 + \MP^2 - m_q^2}
\end{equation}
or, in other words,
\begin{equation}
\label{eq:eff_op}
{\cal O}^{(5)}_{m} =  \frac{1}{2}\frac{y_f^2 m_q}{M_S^2 + \MP^2 - m_q^2} S^2 \bar q q\;
\end{equation}
as in Fig.~\ref{fig:SStoqq}. We keep the quark mass in the denominator, but assume that the DM particles interact at rest.

A well-know consequence of the above is that the cross sections relevant for thermal freeze-out and for indirect detection will differ in general. In particular, in the chiral limit, $m_q \ll M_S$, the LO cross section is  suppressed if $v \ll 1$. This suppression may however be alleviated taking into account radiative corrections~\cite{Bergstrom:1989jr,Flores:1989ru}.

\subsection{First look at internal bremsstrahlung}

\begin{figure}[t]
\centering
\begin{subfigure}[b]{0.2\textwidth}
\includegraphics[width=\textwidth]{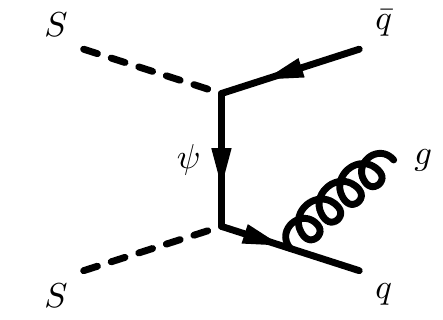}
\subcaption{}
\label{fig:FSRq}
\end{subfigure}
\qquad \qquad
\begin{subfigure}[b]{0.2\textwidth}
\includegraphics[width=\textwidth]{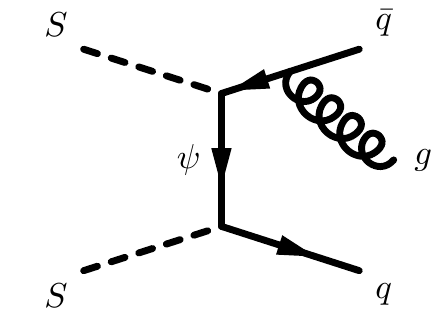}
\subcaption{}
\label{fig:FSRaq}
\end{subfigure}
\qquad \qquad
\begin{subfigure}[b]{0.2\textwidth}
\includegraphics[width=\textwidth]{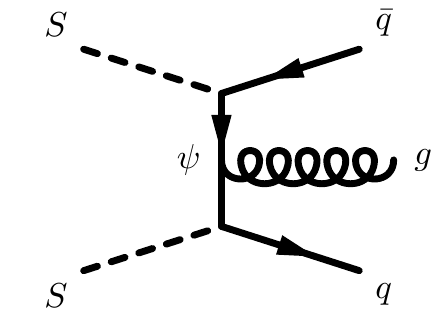}
\subcaption{}
\label{fig:VIB}
\end{subfigure}
\caption{\label{fig:FSR_VIB} Amplitudes contributing to the 3-body process $SS \to q\bar{q} g$. In the body of the text, we refer loosely to final state radiation (FSR) for \ref{fig:FSRq} and \ref{fig:FSRaq} and to virtual internal bremsstrahlung (VIB) for \ref{fig:VIB}.}
\end{figure}

We will focus in this section on QCD corrections (i.e. emission of gluons). Provided $C_F \alpha_s \rightarrow \,Q^2 \alpha$ where $C_F = 4/3$ and $Q$ is the SM fermion electric charge, (most of) our results can be applied to radiation of a gamma instead of a gluon. Now, a pair of $S$ in an s-wave can annihilate into a pair of gluons at one-loop or through internal bremsstrahlung, a 3-body process shown in Fig.~\ref{fig:FSR_VIB}. Although suppressed by powers of $\alpha_s$ or phase-space, these radiative processes may play an important role, both for indirect detection and for setting the relic abundance \cite{Giacchino:2015hvk,Bringmann:2015cpa}.
Annihilation into two gluons has been studied in details in \cite{Ibarra:2014qma,Giacchino:2014moa}, and this for an arbitrary quark mass. Here we focus on internal bremsstrahlung, taking into account quark mass effects. 

The relevant amplitudes are depicted in Figs.~\ref{fig:FSRq} to \ref{fig:VIB}. We will refer loosely to Fig.~\ref{fig:FSRq} and \ref{fig:FSRaq} as final state radiation (FSR) and to Fig.~\ref{fig:VIB} as virtual internal bremsstrahlung (VIB) respectively. 
Then, taking the $S$ particles to be at rest, the amplitude for $S(k_1)S(k_2) \to q(p_1) \bar{q}(p_2) g(k) $ associated to the VIB diagram is given by
\begin{align} \label{eq:ampVIB}
\Mvib =g_s\, y_f^2\,\usp t^a&\Big{\{}P_L\left[ 2(M_S^2+\MP^2-m_q^2)\epsl -\itemmul{\epsilon^*}{(p_1+p_2)} \ksl\right]+ \\ \nonumber
&+m_q\left[\itemmul{\epsilon^*}{(p_1-p_2)}+\epsl \ksl\right]\Big{\}} \vsp D_1 D_2 \;,\nonumber
\end{align}   
where 
\be
D_{i}=\frac{i}{(p_i-K)^2-\MP^2} \;,
\ee
and $K = k_1 = k_2 \equiv (M_S,0)$. 
In this expression, $\epst$ is a shorthand for the polarization vector  $\epsilon^*(k)$ and $t^a$ are the representation matrices for the fundamental of $SU(3)$.

 The FSR amplitudes in Fig.~\ref{fig:FSRq} and \ref{fig:FSRaq} read altogether
\begin{align} \label{eq:ampFSR}
\Mfsr= g_s\, y_f^2\,\usp t^a&\Big{\{} 2m_q \itemmul{\epsilon^*}{(p_1 D_{1k}D_2 -p_2 D_{2k} D_1)}+m_q \epsl \ksl(D_{1k}D_2 +D_1 D_{2k})+ \\ \nonumber
&+2 P_L \epsl \big{[}M_S^2 - M_{\psi}^2 +m_q^2 -\itemmul{K }{(p_1 + p_2)}\big{]}D_1D_2\Big{\}}\vsp \;,
\end{align}
where
\be
D_{ik}=\frac{i}{(p_i-k)^2-m_q^2}\;.
\ee
The last term in Eq.~\eqref{eq:ampFSR} is perhaps surprising, as one could have expected the combination of propagators $D_1 D_2$ to arise only from the VIB amplitude, see Eq.~\eqref{eq:ampVIB}. Concretely, this term comes from the combination
\be
D_1+D_2=-\left[2 \MP^2-2 M_S^2-2m_q^2+2 \itemmul{K}{(p_1+p_2)}\right]D_1 D_2 \,,
\ee
together with $D_{ik}^{-1}= 2 i  \,\itemmul{p_i}{k}$. Actually, this term (which, incidentally, does not vanish in the limit $m_q \rightarrow 0$) is gauge dependent and so must compensate terms from Eq.~\eqref{eq:ampVIB} (notice that this implies that our distinction between FSR and VIB is not clear-cut).
The total amplitude $\Mib  = \varepsilon_\mu \Mib^\mu$ reads
\begin{align} \label{eq:IBamp}
&\Mib =g_s \,y_f^2\, \usp t^a\Big{\{} P_L \left[ \itemmul{(p_1+p_2)}{k} \epsl - \itemmul{\epsilon^*}{(p_1+ p_2)} \ksl \right] D_1 D_2+ \\ \nonumber
&+m_q \Big{[} \itemmul{\epsilon^*}{\left(p_1 (2 D_{1k} +D_1)D_2 - p_2 (2 D_{2k}+D_2)D_1\right)} + \epsl \ksl (D_{1}D_{2k}+D_{2}D_{1k}+D_1 D_2)  \Big{]} \Big{\}} \vsp \;.
\end{align}
 Using 
\be
D_1 - D_2 = -2 i  \,D_1 D_2 \,\itemmul{K}{(p_1 - p_2)} \;,
\ee
one verifies that the total amplitude is gauge invariant, $k_{\mu}\Mib^\mu=0$. 

\bigskip
To gain further insight, it may be instructive to look at Eq.~\eqref{eq:IBamp} from an effective interaction perspective. To do so, we consider an expansion of Eq.~\eqref{eq:IBamp} in $r^{-1} = (M_S/\MP)^2$ assuming $\MP \gg M_S$. Keeping only the dominant contributions, we get the following three terms, each of which is gauge invariant,
\begin{align} \label{eq:AmpDelta1}
\Mib \approx &-\frac{2 g_s y_f^2}{r}\frac{m_q}{M_S^2}\; \usp  t^a\vsp\; \Ieik+ \\ 
\label{eq:AmpDelta2}
& -\frac{g_s y_f^2}{r}\frac{m_q}{M_S^2}\; \usp t^a\epsl \ksl  \left( D_{1k} + D_{2k}\right)\vsp+\\
\label{eq:AmpDelta3}
&+\frac{g_s y_f^2}{r^2}\frac{1}{M_S^4} \; \usp t^a P_L \left[ \itemmul{(p_1+p_2)}{k} \epsl - \itemmul{\epsilon^*}{(p_1+ p_2)} \ksl \right] \vsp \,.
\end{align}
The first two terms are $\propto m_q$. While they cannot be written in terms of local effective operators, they have a simple structure. 
The first term contains the familiar Weizs\" acker-Williams eikonal factor 
$$\Ieik = \frac{\itemmul{\epst}{p_1}}{\itemmul{k}{p_1}} - \frac{\itemmul{\epst}{p_2}}{\itemmul{k}{p_2}},$$
 which multiplies the LO amplitude for $S S \rightarrow q \bar q$ and captures the IR divergences of the total annihilation cross section. 
The second term is IR finite and its numerator has the structure of a dipole interaction, 
\be \label{eq:op7}
\mathcal{O}_{\rm DI} \sim \overline{q}_R  \,\sigma_{\mu \nu} F^{\mu \nu} q_R \;,
\ee 
with $F^{\mu \nu} = t^a F^{a,\mu \nu}$ and $\sigma_{\mu\nu} = i [\gamma_\mu,\gamma_\nu]/2$.
\begin{figure}[t]
\centering
\begin{subfigure}[b]{0.2\textwidth}
\includegraphics[width=\textwidth]{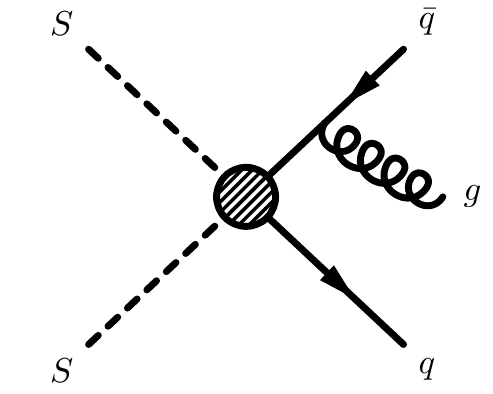}
\subcaption{\label{fig:3B-EFTa}}
\end{subfigure}\qquad \qquad
\begin{subfigure}[b]{0.2\textwidth}
\includegraphics[width=\textwidth]{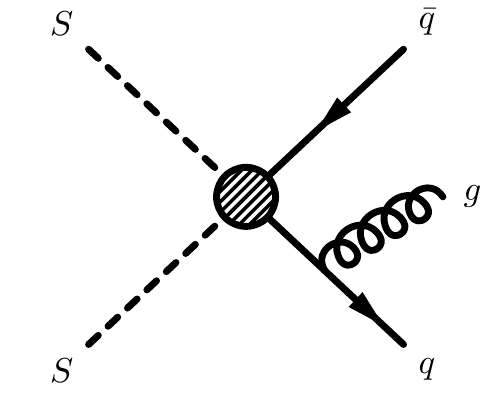}
\subcaption{\label{fig:3B-EFTb}}
\end{subfigure}\qquad \qquad
\begin{subfigure}[b]{0.2\textwidth}
{\includegraphics[width=\textwidth]{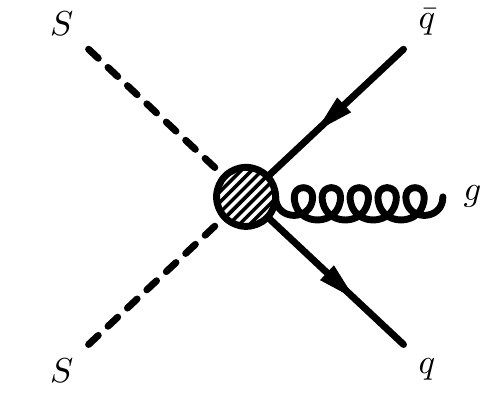}}
\subcaption{\label{fig:3B-EFTc}}
\end{subfigure}
\caption{\label{fig:3B-EFT} Diagrammatic representation of the amplitudes in Eq.~\eqref{eq:AmpDelta1}. We refer to the first two amplitudes, (a) and (b), as FSR, while (c) corresponds to VIB. }
\end{figure}
Due to cancellations between contributions from both $\Mvib$ and $\Mfsr$ amplitudes, it can be seen that this term comes entirely from the $\Mfsr$ amplitude in Eq.~\eqref{eq:ampFSR}. Both contributions involve a chirality flip, $\propto m_q$, like the leading order s-wave annihilation amplitude. These contributions are collectively depicted by the diagrams (a) and (b) in Fig.~\ref{fig:3B-EFT} and we refer to them as FSR amplitudes. Finally, the third term in Eq.~\eqref{eq:AmpDelta1} is local and can be derived from the following dimension eight operator (see diagram (c) in Fig.~\ref{fig:3B-EFT})
\be
\label{eq:VIBeff}
\mathcal{O}_{\rm VIB}^{(8)}= S^2 \partial_{\mu} \left( \overline{q}_R \gamma_{\nu} F^{\mu \nu} q_R\right), 
\ee
already introduced in~\cite{Barger:2009xe,Barger:2011jg}. This term comes from both the $\Mvib$ and $\Mfsr$ amplitudes, but we call it VIB for short, as it reduces  to it  in the chiral limit $m_q/M_S \rightarrow 0$. Incidentally, as it has no helicity suppression, it may be the dominant contribution to $S S$ annihilation if $m_q \ll M_S$. 
In the limit $\MP \gg M_S$ the situation is clear and simple. Concretely, one could use the amplitude of Eq.~\eqref{eq:AmpDelta1} to compute the annihilation cross section. The first term leads to IR divergences, but these can be tamed in the usual way, as we will see below. 
However, here we would like to be more general, first because the large $\MP/M_S$ expansion spoils the spectral feature of VIB, which are most prominent when $\MP$ and $M_S$ are almost degenerate, and second because we have in mind candidates that could annihilate into heavy quarks, in particular the top, so that neglecting $m_q$ may not be a good approximation. 

\begin{figure}[t]
\includegraphics[width=9cm]{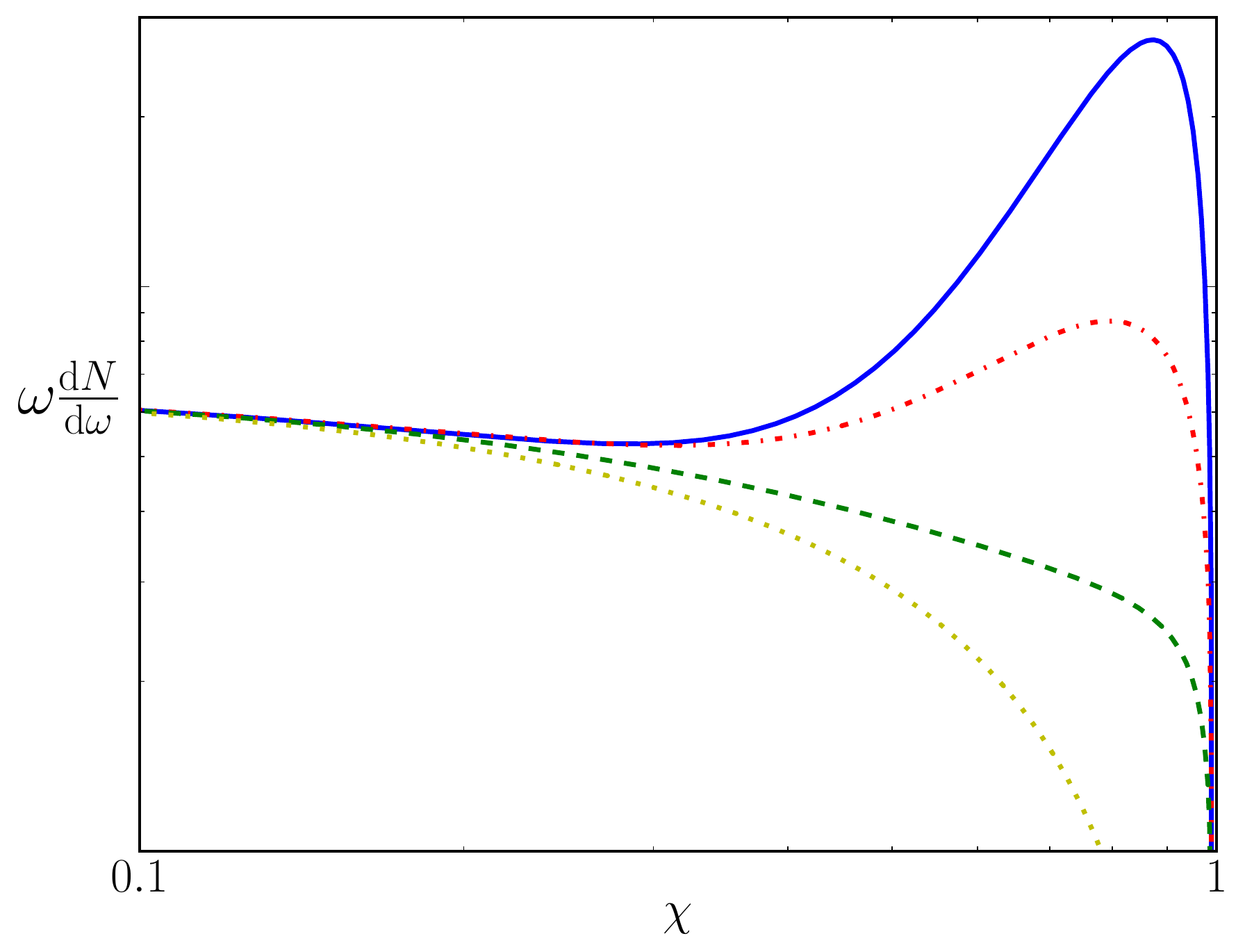}
\caption{\label{fig:InterpolationDiffxSL} In this figure $\omega$ is the gamma or gluon energy and $\chi = \omega/M_s$. It illustrates the behavior of the full differential spectrum at partonic level (blue, solid) compared to the one with only soft emission, following the  Weizs\" acker-Williams approximation Eq.~\eqref{eq:AmpDelta1} (yellow, dotted), together with hard emission Eq.~\eqref{eq:AmpDelta2} (green, dashed) and finally adding the effective VIB contribution of Eq.~\eqref{eq:AmpDelta3} (red, dot-dashed). {Specifically, this figure is for $M_S = 2$ TeV, $r=1.2^2$ and $m_q = m_{\rm top} = 173.5 $ GeV, values for which both the VIB feature and the departure from predictions based on the effective operator of Eq.~\eqref{eq:AmpDelta3}  are clearly visible. The normalization is arbitrary.} 
} 
\end{figure}

Anticipating on the results of the next sections, these  considerations  are illustrated in Fig.~\ref{fig:InterpolationDiffxSL} where we depict the typical gluon or gamma-ray spectrum (at the partonic level) $\omega dN/d\omega$  as function of $\chi = \omega/M_S$ for a DM candidate with a strong VIB feature, thus for almost degenerate masses $M_\psi \gtrsim M_S$, but also a substantial contribution from FSR. The full spectrum, obtained from the amplitudes of Fig.~\ref{fig:FSR_VIB}, is shown as the solid (blue) line. The VIB feature is the peak near $\omega \lesssim M_S$. Emission of soft bosons, corresponding to the Weizs\" acker-Williams approximation, is shown as a dotted (yellow) line. As expected, it captures the behavior of the full spectrum at low energies $\omega \ll M_S$. The two other curves correspond to spectra obtained by using the amplitudes of the
effective theory. The dashed (green) curve is obtained from the amplitudes of Figs.~\ref{fig:3B-EFTa} and \ref{fig:3B-EFTb}. Compared to the Weizs\" acker-Williams approximation, it includes the emission of hard photons or gluons. The spectrum has a sharp edge feature, which is characteristic of FSR \cite{Birkedal:2004xn}. Finally, the dot-dashed (red) curve encompasses all the effective amplitudes of Eqs.~\eqref{eq:AmpDelta1}-\eqref{eq:AmpDelta3}.  While the Weizs\" acker-Williams  approximation reproduces well the emission of soft gluons or gammas, the effective operator in Eq.~\eqref{eq:VIBeff}, corresponding to the amplitude of Eq.~\eqref{eq:AmpDelta3}, fails to fully reproduce the VIB spectral feature.

\subsection{Radiative corrections}

\begin{figure}[t]
\centering
\begin{subfigure}[b]{0.2\textwidth}
\includegraphics[width=\textwidth]{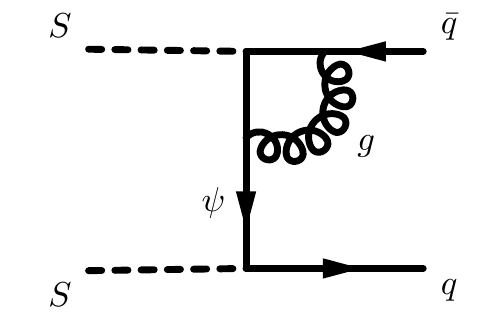}
\end{subfigure}\qquad \qquad
\begin{subfigure}[b]{0.2\textwidth}
\includegraphics[width=\textwidth]{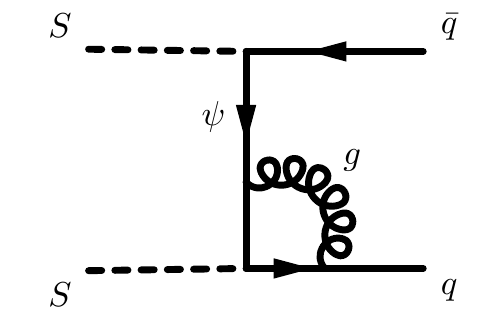}
\end{subfigure}\qquad \qquad
\begin{subfigure}[b]{0.2\textwidth}
{\includegraphics[width=\textwidth]{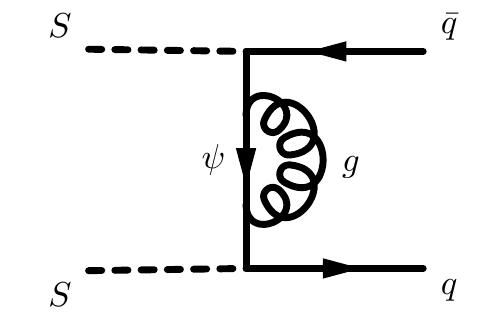}}
\end{subfigure}
\\\vspace{1cm}
\begin{subfigure}[b]{0.2\textwidth}
\includegraphics[width=\textwidth]{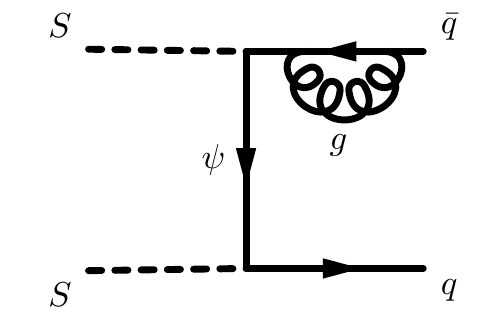}
\end{subfigure}\qquad \qquad
\begin{subfigure}[b]{0.2\textwidth}
\includegraphics[width=\textwidth]{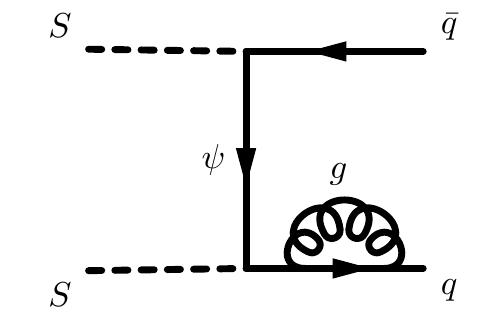}
\end{subfigure}\qquad \qquad
\begin{subfigure}[b]{0.2\textwidth}
{\includegraphics[width=\textwidth]{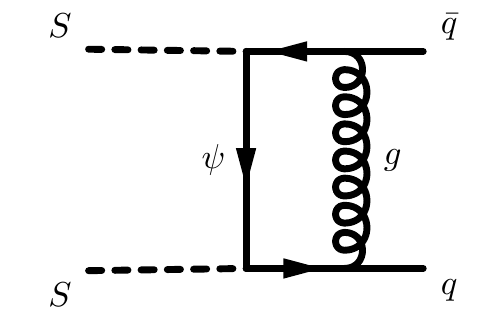}}
\end{subfigure}
\caption{\label{fig:VirtualCorrectionsFull} Full set of one-loop corrections to $SS$ annihilation into $q\bar q$.}
\end{figure}

For the purpose of probing DM through indirect detection, we aim at determining the spectrum of quark and gluons
emitted when DM annihilates through internal bremsstrahlung, 
$
{d \sigma v_{\qqg}}/{d \omega}
$
where $\omega$ is the gluon energy. 
 The integrated cross section is also relevant for determining the relic abundance of the DM particle \cite{Giacchino:2015hvk}. However, for finite quark mass, its expression suffers from IR and collinear divergences.
 The recipe to address  these divergences is standard and involves computing not only the 3-body process, but also the one-loop corrections to the 2-body annihilation. Then, the Kinoshita-Lee-Nauenberg theorem (or Bloch-Nordsieck for QED) states that, order by order in the gauge couplings, IR divergences from phase-space integration, which in our case are ${\cal O}(\alpha_s)$, are cancelled by those from loop corrections. Thus, in principle, we should compute the one-loop corrections depicted in Fig.~\ref{fig:VirtualCorrectionsFull}. 
While such calculations have been performed for minimal supersymmetric candidates (see {\em e.g.} \cite{Herrmann:2014kma}), it is another matter to do so for a  simplified  model. Instead, inspired by the strategy of \cite{Bringmann:2015cpa}, we will separate the problem into the emission of soft and hard gluons. For emission of soft gluons,  we will use {the effective interaction of Eq.(\ref{eq:eff_op})} to control and cancel the IR divergences that affect the cross section for soft modes, while keeping as much as possible the full, UV complete amplitudes to capture the VIB spectral features. {This strategy rests on the fact that, while they differ in the regime of emission of hard gluons or gammas,  both the full theory and the effective interaction of Eq.(\ref{eq:eff_op}) have precisely the same behavior in the IR and in particular lead to the same IR divergent behavior.} We will control the matching between these two regimes using a cut-off on the energy of the emitted gluon, $\omega_0$. The total NLO cross section will take the form
\begin{equation}
\sigma v_{\text{NLO}} = \sigma v_{\text{LO}} + \Delta \sigma v \vert_{\text{soft}}^{\text{eff}}(\omega_0) + \Delta \sigma v \vert_{\text{hard}}^{\text{full}}(\omega_0)\;,
\end{equation}
where $\sigma v_{\text{LO}}  \equiv \sigma v_{q\bar q}$. Both $\Delta \sigma v \vert_{\text{soft}}$ and $\Delta \sigma v \vert_{\text{hard}}$ depend on the matching energy $\omega_0$ but their sum does not.  
In this expression
\begin{equation}
\Delta \sigma v \vert_{\text{soft}}^{\text{eff}}(\omega_0) \equiv \Delta \tilde \sigma v \vert_{\text{soft}}^{\text{eff}}(\omega_0,\lambda) + \Delta \sigma v \vert_{\text{1-loop}}^{\text{eff}}(\lambda) \;,
\end{equation}
where $\lambda$ is a fictitious mass of the gluon, introduced to regularize the cross section obtained by integrating over soft modes. The tilde on $\tilde \sigma$ is there to mean that this cross section is unphysical, as it diverges for $\lambda \rightarrow 0$. As usual, the $\lambda$ dependence requires to take into account one-loop corrections to the LO cross section, which are computed using the effective theory. The $\lambda$ dependence will cancel in the sum of the two contributions. 
Clearly, the main advantage of this  down-to-earth approach is that we will only need to calculate the one-loop corrections depicted in Fig.~\ref{fig:VirtualCorrectionsEFT} to cancel infrared divergences. Incidentally, as this coupling has precisely the same structure as the Higgs coupling to SM fermions, much of the underlying physics is the same as that discussed in  \cite{Drees:1990dq,Braaten:1980yq}. What is specific to the DM scenario is the  emission of gluons by the vector-like mediator.

\subsubsection{Soft gluon emission}

We first consider the annihilation into a pair of massive SM quarks with the emission of a soft gluon. By this, we mean a real gluon with energy $\omega = \vert \vec k\vert \leq \omega_0$, where $\omega_0$ is a cut-off energy, which we take to be small compared to other characteristic mass scale in the theory, $\omega_0 \ll \{M_S,\MP,m_q\}$, but larger than $\Lambda_{\rm QCD}$. In that limit, we describe the emission of a soft gluon using the Weizs\" acker-Williams approximation,
\begin{equation}
{\cal M}^a \vert_{\text{soft}}^{\text{eff}} = - g_s y_f^2\, \frac{m_q}{M_S^2 \Delta }\; \usp  t^a\vsp\;\left(\frac{\itemmul{\epsilon^{\ast}}{p_1}}{\itemmul{k}{p_1}} - \frac{\itemmul{\epsilon^{\ast}}{p_2}}{\itemmul{k}{p_2}}\right)\;.
\end{equation}
This differs from the first term in Eq.~\eqref{eq:AmpDelta1} by the factor $1/\Delta$ with $\Delta  = 1+ \MP^2/M_S^2 - m_q^2/M_S^2 \equiv 1 + r -z$, which stems from neglecting the soft gluon 4-momentum in the propagator of the mediator. 

Integrating over phase space for final state fermions, we get the following differential cross section, valid for emission of a soft gluon of energy $\omega$, 
\begin{eqnarray}
 \label{eq:dSVdEsoft}
\left.\frac{d \sigma v_{\qqg}}{d \chi}\right\vert^{\text{eff}}_{\text{soft}}&=& \frac{y_f^4 N_c}{4\pi \Delta^2 M_S^2}\frac{\alpha_S C_F}{\pi} \left\{ 
\frac{\left(2-z\right)\left(1-z\right)z}{\chi}\log\frac{\chi+\beta\sqrt{\chi^2-4\mu}}{\chi-\beta \sqrt{\chi^2-4\mu}}+\right.
\nonumber\\
& & \left. - 2\left(1-z\right)z^2 \frac{\beta \sqrt{\chi^2-4\mu}}{\left(1-\beta^2\right)\chi^2+4\beta^2\mu}\right\}\;,
\end{eqnarray}
where $C_F =4/3$ and $\chi = \omega/M_S$. This expression involves the velocity of the final state quarks in the rest frame of the $q\overline{q}$ system (see {\em e.g.} \cite{DeWit:1986it}) 
\be
\beta=\sqrt{\frac{1- \chi  -z + \mu}{1-\chi+ \mu}}\quad \xrightarrow[\omega,  \lambda\to 0] \qquad \beta_0 = \sqrt{1-z} \equiv \sqrt{ 1- \frac{m_q^2}{M_S^2}}\;.
\ee
 To regulate the IR divergence that will arise when integrating Eq.~(\ref{eq:dSVdEsoft}) over the gluon energy, we have introduced a fictitious mass $\lambda \ll \omega_0$ for the soft gluon, which appears through $\mu = \lambda^2/4 M_S^2$.
 
Keeping only the leading terms in the limit $\lambda\rightarrow 0$, the integrated cross section for emission of a soft gluon in the energy range $\lambda \leq \omega\leq  \omega_0$ is given by
\begin{eqnarray}
\label{eq:soft}
\Delta \tilde \sigma  v\vert_{\text{soft}}^{\text{eff}}&=&\sigma v_{q \bar q}\, \frac{\alpha_s C_F}{\pi}\left\{\left(\frac{1+ \beta_0^2}{2 \beta_0}\log \frac{1+\beta_0}{1-\beta_0}-1\right)\log \frac{4\omega_0^2}{\lambda^2}+\right. \\
&+& \frac{1+\beta_0^2}{\beta_0}\left[\text{Li}_2\left(\frac{1-\beta_0}{1+\beta_0}\right)+\log\frac{1+\beta_0}{2 \beta_0}\log \frac{1+\beta_0}{1-\beta_0}+\right.\nonumber\\
&-& \left.\frac{1}{4}\log^2\left(\frac{1+\beta_0}{1-\beta_0}\right)-\frac{\pi^2}{6} \right] 
 + \left. \frac{1}{\beta_0}\log \frac{1+\beta_0}{1-\beta_0}\right\}.\nonumber
\end{eqnarray}
By construction, this cross section is proportional to the leading order cross section $\sigma v_{q \bar q}$, which corresponds here to the s-wave part of Eq.~(\ref{eq:SStoqq}).
This expression can be compared (and agrees) with that of \cite{Drees:1990dq} for decay of the Higgs, with which it shares the IR divergence term $\propto \log(\omega_0^2/\lambda^2)$. 

\subsubsection{Virtual one-loop corrections} \label{sec:Virtual1L}
\begin{figure}[t]
\centering
\begin{subfigure}[b]{0.2\textwidth}
\includegraphics[width=\textwidth]{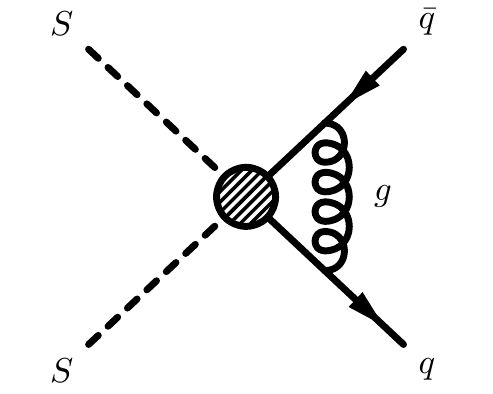}
\end{subfigure}\qquad \qquad
\begin{subfigure}[b]{0.2\textwidth}
\includegraphics[width=\textwidth]{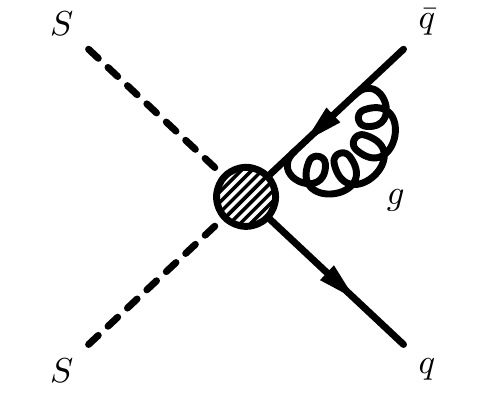}
\end{subfigure}\qquad \qquad
\begin{subfigure}[b]{0.2\textwidth}
{\includegraphics[width=\textwidth]{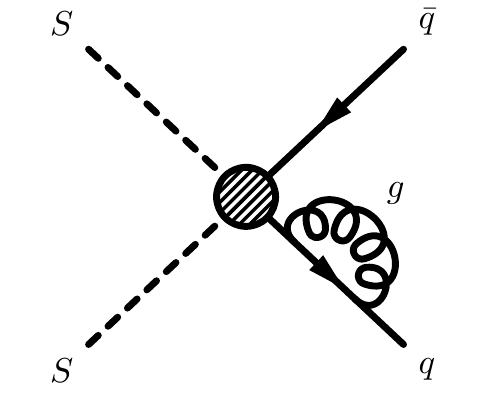}}
\end{subfigure}
\caption{\label{fig:VirtualCorrectionsEFT}One-loop corrections to the effective coupling $SS\to q\bar{q}$ relevant for cancelling IR divergences (see Sec.~\ref{sec:Virtual1L}).}
\end{figure}
Physical cross sections should be free of IR divergences as $\lambda\rightarrow 0$. To obtain a IR finite result, we need take into account the contributions of ${\cal O}(\alpha_s)$ virtual one-loop corrections to the leading order cross section into $q\bar q$. This stems from the interference term between LO and the one-loop corrections
$$
\vert {\cal M}_{\rm tree} + {\cal M}_{\rm 1-loop}\vert^2 = \vert {\cal M}_{\rm tree}\vert^2 + 2\, {\rm Re}({\cal M}_{\rm tree}^\ast {\cal M}_{\rm 1-loop}) + {\cal O}(\alpha_s^2) \;.
$$
At one-loop, the IR divergent contributions come from the vertex correction and final state fermion wave-function corrections, depicted by the diagrams of Fig. \ref{fig:VirtualCorrectionsEFT}.

Using dimensional regularisation in $D=4 - 2 \epsilon$, the virtual correction to the effective vertex is given by (see \cite{Drees:1990dq} for comparison)
\begin{eqnarray}
{\rm Re} \mathcal{M}\vert_{\text{1-loop}}^{\text{eff}}&=&\mathcal{M}_{\text{tree}}\frac{\alpha_s C_F}{2 \pi}\left[2\left(\frac{1}{\epsilon}-\log \frac{m_q^2}{\mu^2}\right)-\frac{1+\beta_0^2}{2\beta_0}\log\frac{1+\beta_0}{1-\beta_0}\log\frac{m_q^2}{\lambda^2}+\right.\nonumber\\
&+&\left.\frac{1+\beta_0^2}{\beta_0}\left\{\text{Li}_2\left(\frac{1-\beta_0}{1+\beta_0}\right)+\log\frac{1+\beta_0}{2\beta_0}\log\frac{1+\beta_0}{1-\beta_0}-\frac{1}{4}\log^2\frac{1+\beta_0}{1-\beta_0}+\frac{\pi^2}{3}\right\}+\right.\nonumber\\
&+&\left.\frac{1-\beta_0^2}{\beta_0}\log\frac{1+\beta_0}{1-\beta_0}+3\right]\;,
\end{eqnarray}
which is both UV and IR divergent.\footnote{{The  LO amplitude ${\cal M}_{\rm tree}$ is real, so we only need the real part of  ${\cal M}_{\rm 1-loop}$, see also  \cite{Drees:1990dq}.  As both amplitudes are $\propto \vert y_f\vert^2$, this is true regardless of the phase of the Yukawa coupling.}}

According to the Lehmann-Symanzik-Zimmermann (LSZ) reduction formula \cite{Peskin:1995ev}, we must also take into account the ${\cal O}(\alpha_s)$ correction from the one-shell wave-function of the final state quark and anti-quark, with \footnote{{The relevance of wave-function renormalization may also be understood as follows. Both the one-loop correction to the effective vertex and the final state fermion wave-function correction ({\em i.e.} $Z_2$) are infrared divergent to ${\cal O}(\alpha_s)$. To the same order, their infrared divergences are cancelled by taking the infrared divergences of FSR from radiation amplitudes. A detail analysis reveals (see {\em e.g.} the lectures notes by  D. Ross \cite{Ross:2018}) that the IR divergence from the correction to the vertex is cancelled by the interference term, that is emission from {distinct legs}, while the one from the $Z_2$ factor is cancelled by the square of each FSR amplitudes,  corresponding to emission from {same legs}.}}
\be
\left(Z_2-1\right) \mathcal{M}_{\text{tree}}=\delta_2 \mathcal{M}_{\text{tree}}  = \mathcal{M}_{\text{tree}}\frac{\alpha_s C_F }{2 \pi}\left[ -{1\over2}\left(\frac{1}{\epsilon} -\log\frac{m_q^2}{\mu^2}\right)+ \log\frac{m_q^2}{\lambda^2} -2\right] \;.
\ee

Thus, the one-loop unsubstracted correction to annihilation into a quark-antiquark pair is
\begin{eqnarray}
\label{eq:oneloop}
{\rm Re}(\mathcal{M}_{\text{1-loop}})&=&\mathcal{M}_{\text{tree}}\frac{\alpha_s C_F}{2 \pi}\left[{3\over 2}\left(\frac{1}{\epsilon}-\log \frac{m_q^2}{\mu^2}\right)-\left(\frac{1+\beta_0^2}{2\beta_0}\log\frac{1+\beta_0}{1-\beta_0}-1\right)\log\frac{m_q^2}{\lambda^2}+\right.\nonumber\\
&+&\left.\frac{1+\beta_0^2}{\beta_0}\left\{\text{Li}_2\left(\frac{1-\beta_0}{1+\beta_0}\right)+\log\frac{1+\beta_0}{2\beta_0}\log\frac{1+\beta_0}{1-\beta_0}-\frac{1}{4}\log^2\frac{1+\beta_0}{1-\beta_0}+\frac{\pi^2}{3}\right\}+\right.\nonumber\\
&+&\left.\frac{1-\beta_0^2}{\beta_0}\log\frac{1+\beta_0}{1-\beta_0}+1\right]\;.
\end{eqnarray}
Comparing the second term of this expression to the first term in Eq.~\eqref{eq:soft}, we see that, adding the ${\cal O}(\alpha_s)$ one-loop corrections to the tree level cross section to the cross section for emission of soft gluons, the dependence on the fictitious gluon mass ({\em i.e.} the terms in $\log(\lambda^2)$) disappears \cite{Drees:1990dq}, leaving only the dependence on the cut-off on the energy of the emitted gluon $\propto \log(\omega_0^2)$. 

The resulting expression has still a UV divergence, which must be  appropriately cancelled. The renormalization prescription used in \cite{Bringmann:2015cpa} is the same as the one advocated in \cite{Drees:1990dq} in the case of QCD corrections to Higgs decay into quarks. 
In this case, since the current quark mass stems from Yukawa coupling to the Higgs, the counter-term is that for quark mass renormalization, 
\be
\label{eq:massren}
\frac{\delta m_q}{m_q}=-\frac{C_F \alpha_s}{2 \pi}\left[ {3\over 2}\left( \frac{1}{\epsilon} - \log\frac{m_q^2}{\mu^2}\right)+2\right]\;.
\ee
This term clearly cancels the UV divergent part of Eq.~\eqref{eq:oneloop} but any other prescription would only differ from this choice by a constant term. Which choice one makes does not matter. Indeed, for fixed particle masses, the only free parameters in the  annihilation cross section (meaning here at ${\cal O}(\alpha_s)$) is the Yukawa coupling $y_f$. Its value is fixed by matching to the cosmic relic abundance. All other parameters being kept fixed, a different renormalization prescription just amounts to fixing $y_f$ to a (slightly) distinct value. For definiteness, here we use the same prescription of \cite{Bringmann:2015cpa} to renormalize our effective theory.\footnote{ 
The correspondence with the problem of QCD corrections to SM Higgs decay into quarks rests on the use of an effective vertex. In principle, a procedure we could follow is to match our effective theory with the more complete theory at the scale(s) at which one integrates out the heavy degree(s) of freedom. For instance,  the one-loop corrections include the box diagram depicted in Fig.~\ref{fig:VirtualCorrectionsFull}, which has a better UV behavior than in the effective theory. With the mass of vector-like quark, $M_\psi^2$ acting as a cut-off, divergent terms $1/\epsilon$ could actually correspond to $\propto \log (M_\psi^2/M_S^2)$ contributions (see e.g. \cite{Peskin:1995ev}). Using the matching procedure would only introduce minor corrections (at least compared to the major impact of taking into account bremsstrahlung). 
See \cite{Bringmann:2015cpa} for further considerations on errors from using the effective approach.}

Doing so, we get the one-loop correction to the LO cross section
\begin{eqnarray}
\label{eq:soft_nlo}
\Delta \sigma v\vert_{\text{1-loop}}^{\text{eff}} 
&=& \sigma v_{q \bar q}\, \frac{\alpha_s C_F}{ \pi}\left[-\left(\frac{1+\beta_0^2}{2\beta_0}\log\frac{1+\beta_0}{1-\beta_0}-1\right)\log\frac{m_q^2}{\lambda^2}+\right.\nonumber\\
&+&\left.\frac{1+\beta_0^2}{\beta_0}\left\{\text{Li}_2\left(\frac{1-\beta_0}{1+\beta_0}\right)+\log\frac{1+\beta_0}{2\beta_0}\log\frac{1+\beta_0}{1-\beta_0}-\frac{1}{4}\log^2\frac{1+\beta_0}{1-\beta_0}+\frac{\pi^2}{3}\right\}+\right.\nonumber\\
&+&\left.\frac{1-\beta_0^2}{\beta_0}\log\frac{1+\beta_0}{1-\beta_0}) - 1\right]\;.
\end{eqnarray}
Adding this contribution to Eq.~\eqref{eq:soft} gives $\Delta \sigma v\vert_{\text{soft}}^{\text{eff}}$, which depends on the cut-off energy $\omega_0$ but not on $\lambda$, {\em i.e.}
\be
\label{eq:soft_eff}
\Delta \sigma v\vert_{\text{soft}}^{\text{eff}} = \Delta \tilde \sigma v\vert_{\text{soft}}^{\text{eff}}+\Delta  \sigma v\vert_{\text{1-loop}}^{\text{eff}}= \sigma v_{q\bar q}\,\frac{\alpha_s C_F}{\pi}\left[\Big(1- \frac{1+\bo^2}{2\bo}\log\frac{1+\bo}{1-\bo}\Big)\log \frac{m_q^2}{\omega_0^2} + \ldots
)
\right]\;,
\ee
where the dots correspond to terms that are ${\cal O}(\omega_0^0)$.

\subsubsection{Hard gluon emission}
It remains to determine the spectrum of hard gluons and their contribution to the total NLO cross section. For this, we use the full theory, including the effects of the vector-like particle, from the amplitudes of Fig.~\ref{fig:FSR_VIB}. Since we will put a cut-off on the energy of the gluon, no gluon mass term is required. The calculations, although cumbersome, are straightforward. The differential cross section can be written as
\begin{eqnarray}
 \label{eq:dSVdEhard}
\left.\frac{d \sigma v_{\qqg}}{d \chi}\right\vert_{\text{full}}&=& \frac{y_f^4N_c }{4 \pi M_S^2}\frac{\alpha_s C_F}{\pi } \left\{ 
\frac{\left(2-z\right)\left(1-z\right)z}{ \Delta^2 \chi}\log\frac{1+\beta}{1-\beta}+\right.
\nonumber\\
& & \left. - {2\left(1-z\right)z^2\over \Delta^2} \frac{\beta}{\left(1-\beta^2\right)}\frac{1}{\chi} +  S_0(\chi)\right\}\;.
\end{eqnarray}
In this expression, we have separated the terms that are divergent in the limit $\chi = \omega/M_S \rightarrow 0$ from those that are regular; the latter are collectively expressed as the function $S_0(\chi)$, whose expression is extraordinarily long and not particularly illuminating; its full expression is given for the sake of reference in Appendix~\ref{app:sv}. It contains in particular contributions that reduce to the known expression of virtual internal bremsstrahlung in the limit of massless quarks, see Eq.~\eqref{eq:VIB0}. For finite quark masses, it also includes hard emission from final state quarks and interference terms between the latter and VIB. 

Integrating Eq.~\eqref{eq:dSVdEhard} over $\omega \geq \omega_0$, we get
\begin{align} \label{eq:sv_hard}
\Delta\sigma v\vert_{\text{hard}}^{\text{full}}&=\sigma v_{q\bar q}\,\frac{\alpha_s C_F}{\pi}\Bigg[-\Big(1- \frac{1+\bo^2}{2\bo}\log\frac{1+\bo}{1-\bo}\Big)\log \frac{\bo^4 M_S^2}{\omega_0^2}+\\
&+ 2 \left(\log \frac{1-\bo^2}{4}
+\frac{1+\bo^2}{2\bo}\log\frac{1+\bo}{1-\bo}+1\right)+\nonumber\\
&+\frac{1+\bo^2}{\bo}\Bigg(2\text{Li}_2\left(\frac{1-\bo}{1+\bo}\right)+2\text{Li}_2\left(-\frac{1-\bo}{1+\bo}\right) -\frac{\pi^2}{6}+2\log\frac{1+\bo}{2\bo}\log\frac{1+\bo}{1-\bo}\Bigg)\Bigg]+ \nonumber\\
&+ \frac{N_C}{4 \pi^2}C_F \frac{\alpha_s y_f^4}{M_S^2}\int\limits_{0}^{\bo^2} d \chi S_0(\chi)\;.\nonumber
\end{align}
The first term in this expression involves the cut-off energy $\omega_0$. Adding $\Delta\sigma v\vert_{\text{hard}}^{\text{full}}$ to 
the soft contribution gives a result that is independent of $\omega_0$. Our final expression for the cross section for s-wave annihilation is then
\begin{eqnarray}
\label{eq:final_tot}
\sigma v_{\text{NLO}} &=& \sigma v_{\text{LO}} + \Delta \sigma v \vert_{\text{soft}}^{\text{eff}}(\omega_0) + \Delta \sigma v \vert_{\text{hard}}^{\text{full}}(\omega_0) \nonumber\\
 &=& \sigma v_{q\bar q}\left\{ 1 + \,\frac{\alpha_s C_F}{\pi}\left[\left(1- \frac{1+\bo^2}{2\bo}\log\frac{1+\bo}{1-\bo}\right)\log \frac{1-\bo^2}{4\bo^4} +\right. \right. \nonumber\\
 &  &+ \frac{1+\bo^2}{\bo}\left(4 \text{Li}_2\left(\frac{1-\bo}{1+\bo}\right)+2\text{Li}_2\left(-\frac{1-\bo}{1+\bo}\right)+4\log\frac{1+\bo}{2\bo}\log\frac{1+\bo}{1-\bo}-\frac{1}{2}\log^2\frac{1+\bo}{1-\bo}\right)+ \nonumber\\
 &  &+ \left. \left. 2\log\frac{1-\bo^2}{4}+\frac{3}{\bo}\log\frac{1+\bo}{1-\bo}+\frac{1}{2} \right]\right\}+\nonumber \\
 &  &+ \frac{\alpha_S C_F}{4\pi^2}\frac{y_f^4 N_c}{M_S^2}\int \limits_0^{\bo^2} d \chi S_0\left(\chi\right).
\end{eqnarray}
which is one of our main results.

\subsection{Discussion}
\label{sec:discussion}

The expression of Eq.~\eqref{eq:final_tot} is free of infrared divergences and thus is {\em a priori} useful to determine the relic abundance of $S$ particles and its indirect signatures. It is however too complex to be practical. Furthermore, and despite its complexity, 
it still has some limitations. First, the expression in Eq.~\eqref{eq:final_tot} diverges close to threshold for quark-antiquark pair production. Second, due to collinear divergences, it is  pathological in the opposite limit, $m_q \ll M_S$, or $\beta_0 \rightarrow 1$. The same problems arise in the calculation of QCD corrections to the hadronic decay of the Higgs and the way to solve them is essentially the same. In this section, we first briefly discuss what we should do in principle for dark matter annihilation, and then we explain what we do in practice (see also Ref.\cite{Bringmann:2015cpa} for a similar discussion).  By the same token, we discuss in this section approximations that can be used to take into account the relevant aspects of radiative corrections to the DM model without resorting to the full complexities of the ${\cal O}(\alpha_s)$ cross section. 

We first dispose of the problem posed by the divergent behavior of the NLO cross section close to threshold for fermion-antifermion production. For $M_S \gtrsim m_q$, corresponding to $\beta_0 = \sqrt{1- m_q^2/M_S^2} \rightarrow 0$, the annihilation cross section behaves as
\be
\left.\sigma v_{\text{NLO}}\right\vert_{M_S \gtrsim m_q}  \approx \sigma v_{{\text{LO}}} \left[1 + \frac{\alpha_s C_F}{\pi} \left({\pi^2\over 2 \beta_0} - 1\right)\right]
\ee
The ${\cal O}(\alpha_s$) terms arise from expanding near zero velocity $\beta_0 \rightarrow 0$ the two Spence functions in Eq.~\eqref{eq:final_tot}. As shown in \cite{Drees:1990dq}, this singular behavior is spurious, as the cross section should be in a p-wave quark-antiquark final state, $\propto \beta_0^3$.  It can be traced entirely to the virtual correction to the effective vertex in Fig.~\ref{fig:VirtualCorrectionsEFT} or, in other words, from Eq.~\eqref{eq:soft_nlo}. Physically it signals the tendency to form a bound state, so in principle one should sum an infinite number of diagrams or, below threshold, take into account a possible quarkonium bound state \cite{Drees:1989du}. We consider this to be beyond our scope. Now, for $b\bar b$ (or $\tau^+\tau^-$) the mass of the dark matter and its charged partners is too small to provide viable DM scenario, {\em i.e.} we  are always far from threshold. A case of potential interest is a top-philic scenario, with annihilation of DM into top-antitop pairs ($gg$) above (resp. below) threshold \cite{Colucci:2018vxz}. As the threshold for top-antitop corresponds to a very specific and narrow region of the parameter space, we may take the LO annihilation cross section as a proxy for the true cross section, $\sigma v_{q\bar q} (1 + K)$,  with a K-factor that we  estimate empirically (see \cite{Colucci:2018vxz}).  

The cross section is also pathological  in the opposite limit $M_S \gg m_q$ or $\beta_0 \rightarrow 1$. In particular, the part that is proportional to $\sigma v_{\rm LO}$ receives a large negative logarithmic contribution~\cite{Braaten:1980yq,Drees:1990dq},
\begin{equation}
\label{eq:div_log}
\sigma v_{\text{NLO}} \approx \sigma v_{q \bar q} \left\{ 1  +\frac{\alpha_s C_F}{ \pi}\left[ {9\over 4}- {3\over 2} \log\left({4 M_S^2\over m_q^2}\right) \right]\right\}+ \ldots
\end{equation}
The dots represent the terms that are regular for small  $z = m_q^2/M_S^2$. Importantly, they reduce to the cross section for the pure VIB process in the limit $z \rightarrow 0$, see Eq.~\eqref{eq:VIB0}, a regime in which $\sigma v_{q \bar q} \rightarrow 0$ so, {\em a priori}, the new logarithmic term is harmless for most of the parameter space. If not, one must in principle re-sum large logarithmic contributions to the cross section. A simple recipe to address this is to notice that, as in the case of the Higgs decay, the leading log term in Eq.~\eqref{eq:div_log} is precisely the one-loop ${\cal O}(\alpha_s)$ correction to the quark mass operator \cite{Bringmann:2015cpa}. 
 A convenient way to have a regularized expression for the NLO cross section is thus to subtract from Eq.~\eqref{eq:final_tot} the logarithmically divergent term in Eq.~\eqref{eq:final_tot} and replacing in the expression for $\sigma v_{q\bar q}$ the quark mass parameter $m_q$ by the running mass \cite{Drees:1990dq}
\begin{equation}
\label{eq:running}
m_q \rightarrow \bar m(M_S) =m_q  \left({\log(m_q^2/\Lambda^2)\over \log(M_S^2/\Lambda^2)}\right)^{4\over b_0}
\end{equation}
with $b_0 = 11 - 2/3\; n_f$ the leading order function with $n_f$ the number of quarks lighter than the scale  $2 M_S$ and $\alpha_s(Q^2)  = 4 \pi/(b_0\log(Q^2/\Lambda^2))$.\footnote{{More precisely, the proposed recipe is to replace the factor $m_q^2$ in Eq.(\ref{eq:SStoqq}) by $m_q \rightarrow \bar m(M_S)$. Indeed, expanding $ \bar m^2(M_S)$ to leading order in $\alpha_s$ gives
$$
\bar m^2(M_S) \approx m_q^2\left(1 - {3\over 2} {C_F\over \pi} \alpha_s \log\left({M_S^2\over m_q^2}\right)\right)
$$
which reproduces  the  term that diverges as $m_q \rightarrow 0$. The  running mass is of course regular as $ m_q \rightarrow 0$. 
}}

\begin{figure}[t]
\centering
\begin{subfigure}[b]{8cm}
\includegraphics[width=\textwidth]{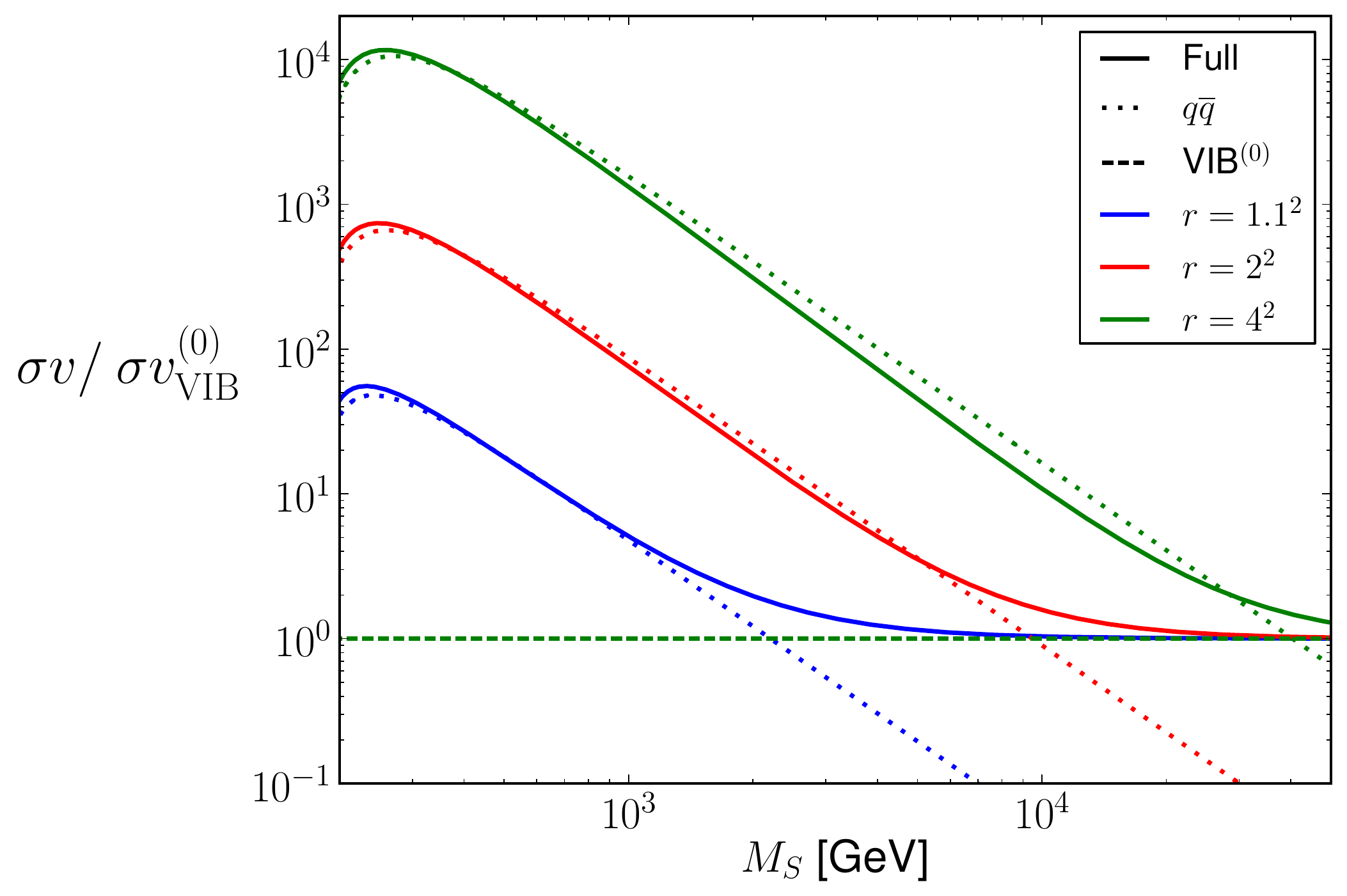}
\subcaption{
\label{fig:InterpolationxSL}}
\end{subfigure}
\quad \quad
\begin{subfigure}[b]{7cm}
\includegraphics[width=\textwidth]{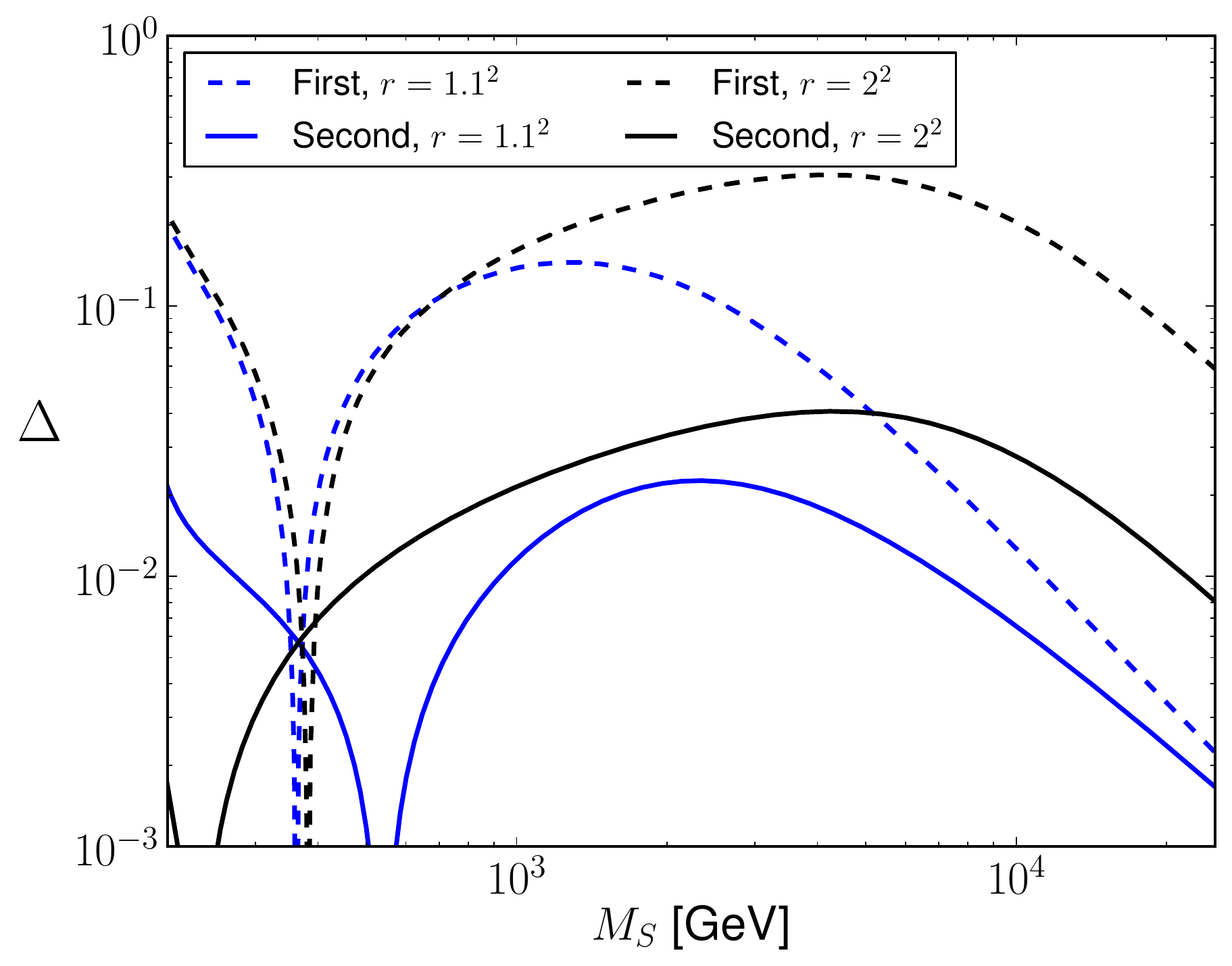}
\subcaption{\label{fig:InterpolationxSR}}
\end{subfigure}
\caption{Left panel: Ratio $\sigma v_{\text{NLO}}/\sigma v_{\text{VIB}}^{(0)}$ as function of the DM mass $M_S$ for three characteristic values of $r = \MP^2/M_S^2$. The curves are shown for $m_q = m_t$, the top quark mass. The dotted lines corresponds to $\sigma v_{q \bar q}$. By definition, the horizontal dashed line corresponds to the VIB cross section in the massless limit, $\sigma v_{\text{VIB}}^{(0)}$. Right panel: Relative errors $\Delta$ due to use of the approximate expression of Eq.~\eqref{eq:InterpolationWithTreeLevel} (First) or Eq.~\eqref{eq:InterpolationBetter} (Second).}
\end{figure}

\bigskip
Now that we have a complete and reliable description of the NLO effects, we go on discussing controllable approximations that can be made to obtain simple and practical expressions for the annihilation cross section with FSR and VIB. In Fig.~\ref{fig:InterpolationxSL} we show (solid lines) the full, NLO total cross section ({\em i.e.} expression in Eq.~\eqref{eq:final_tot}, modulo the caveats discussed above) as function of the DM mass $M_S$ for three benchmark values of the mass ratio $r = (\MP/M_S)^2 = \{1.0, 2^2,4^2\}$ and for $z = (m_t/M_S)^2$ with $m_t$ the top quark mass. Concretely, what we show is the ratio $\sigma v_{\text{NLO}}/\sigma v_{\text{VIB}}^{(0)}$ where $\sigma v_{\text{VIB}}^{(0)}$ is the VIB cross section in the limit of zero quark mass, Eq.~\eqref{eq:VIB0}. Clearly, at large DM mass $M_S \gg m_q$, $\sigma v_{\text{NLO}} \rightarrow \sigma v_{\text{VIB}}^{(0)}$. For lower DM masses, the cross section is dominated by the chirally suppressed component $\propto \sigma v_{q \bar q}$. For increasing $r =(\MP/M_S)^2$, the VIB contribution is relatively suppressed, as it scales like $\sigma v_{\text{VIB}} \propto r^{-4}$ while $\sigma v_{q \bar q}  \propto r^{-2}$, see Eq.~\eqref{eq:AmpDelta1}. The cross-over between the two regimes may be read from Fig.~\ref{fig:InterpolationxSL} where the dotted lines correspond to the leading order $\sigma v_{q \bar q}$ cross section. Crossing of $\sigma v_{q \bar q}$ and the VIB cross section occurs roughly for 
$M_S/m_q \approx r \, \sqrt{2 \pi/0.21 C_F \alpha_s}$ with $m_q \equiv m_t$ in the figure and the factor $0.21 \equiv 7/2 - \pi^2/6$, see Eq.~\eqref{eq:zeroVIB}. 

The main lesson is that the NLO cross section $\sigma v^{\text{NLO}} $ is  reasonably approximated by the following simple expression, in which the leading 2-body cross section $\sigma v_{q \bar q}$ is added to the VIB   cross section in the massless limit, 
\begin{equation}
\label{eq:InterpolationWithTreeLevel}
\sigma v_{\text{NLO}} 
\approx \sigma v_{q\overline{q}}+ \sigma v_{\text{VIB}}^{(0)} \;.
\end{equation}
Doing so, the relative error ({\em i.e.} the K-factor from QCD corrections)  is ${\cal O}(20 \%)$, see Fig.~\ref{fig:InterpolationxSR},  
If necessary, an even better approximation, good to within a few percents (see solid lines in Fig.~\ref{fig:InterpolationxSR}), is obtained by replacing $\sigma v_{q \bar q}$ with the NLO expression in the effective theory, see Eq.~\eqref{eq:drees_NLO}, or 
\begin{equation}
\label{eq:InterpolationBetter}
\sigma v_{\text{NLO}} 
\approx \sigma v_{q\overline{q}}^{\text{NLO}}+ \sigma v_{\text{VIB}}^{(0)}\;.
\end{equation}
This is little surprise, as our NLO calculation is built upon the effective operator in Eq.~\eqref{eq:LOsoft}, which should lead to the dominant contribution to the cross section when VIB emission may be neglected. Our calculations show that interference terms play little role, even when VIB is relevant. More difficult to assess is the error made by using the effective theory instead of the full one-loop amplitudes depicted in Fig.~\ref{fig:VirtualCorrectionsFull}, but it should not be more than a few percent, based on the experienced gained in the case of Majorana dark matter \cite{Bringmann:2015cpa}. In \cite{Colucci:2018vxz} we have used the approximation of 
Eq.~\eqref{eq:InterpolationWithTreeLevel}, which is easy to implement in numerical codes for DM abundance calculations, like  {\sc MicrOMEGAs}~\cite{Belanger:2013oya}, just adding the massless 3-body process associated to VIB, as in Eq.~\eqref{eq:zeroVIB}, when it is relevant.

\section{Differential cross sections and gamma-ray spectra}
\label{sec:diff}

\subsection{Differential cross sections}
\label{sec:approximations}
\begin{figure}[t]
\includegraphics[width=8cm]{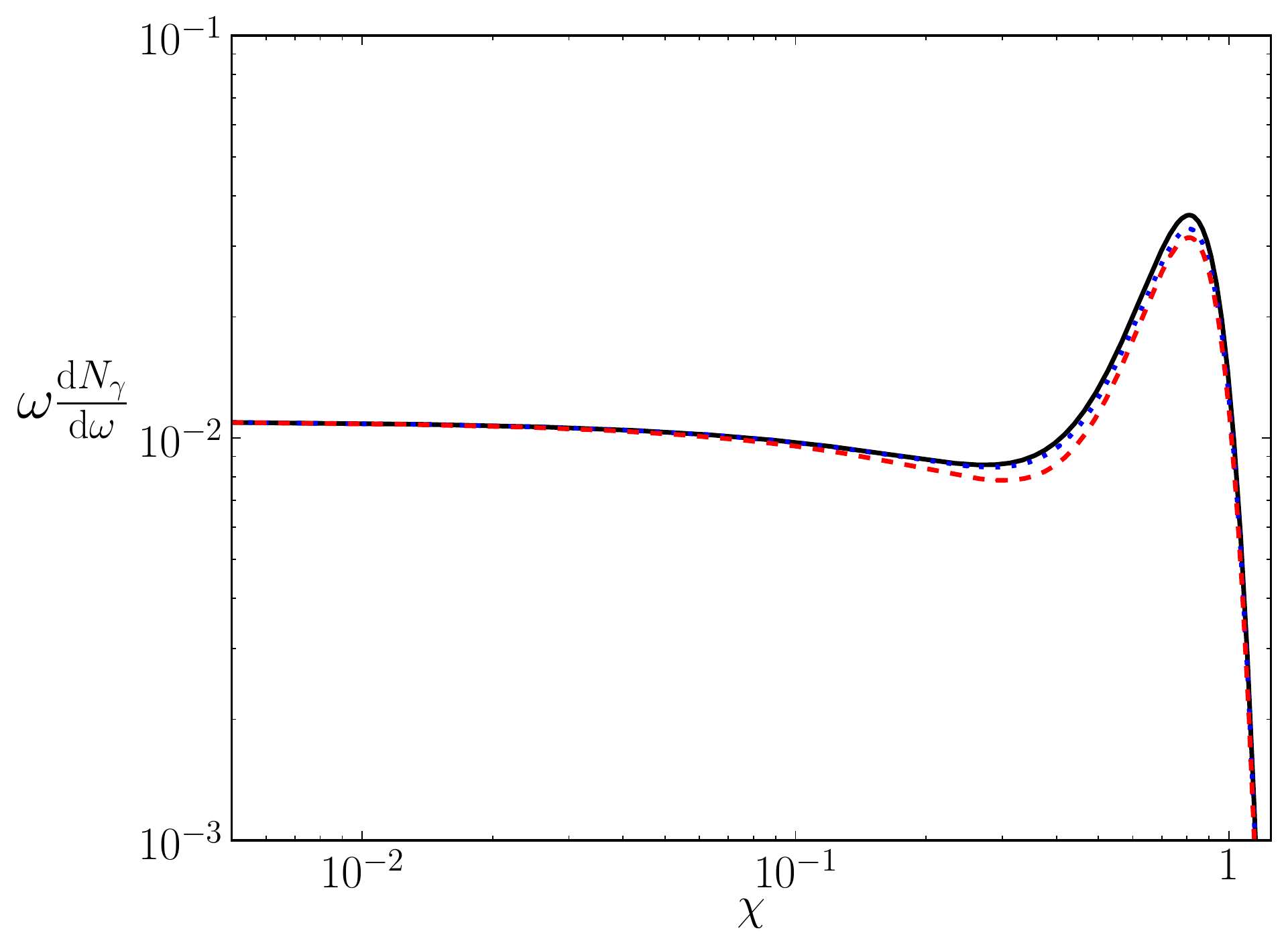}
\caption{\label{fig:InterpolationDiffxSR} Gamma-ray spectra (partonic level) from an $M_S = 2$ TeV DM candidate annihilating into top-antitop using different approximations. The vector-like mediator has a mass $\MP = 2.4$ TeV.  The spectra has been normalized to the LO 2-body cross section. The black solid line corresponds to the full expression. The blue dotted one to the approximation of Eq.~\eqref{eq:diff_approx1} and the dashed red line to the approximation of Eq.~\eqref{eq:split}. The spectra have been convolved with a Gaussian window function with resolution of $10\%$.} 
\end{figure}

The total annihilation cross section is not only relevant for determining the relic abundance; it also sets the scale for indirect searches. But for the latter purpose, we need an handle on the differential cross sections, both in gamma-rays and in gluons, so as to build their spectra, 
$$
 {dN\over d \omega}  = {1\over \sigma} {d \sigma\over d\omega}\;,
 $$
where here $\omega$ stands for the energy of the emitted gamma or gluon and $\sigma$ is the total  annihilation cross section into this channel. This is {\em a priori} straightforward but in practice, things are more complicated, as to assess the indirect signature from DM annihilation, say into gamma-rays, what we need is to take into account both the contribution of gamma-ray produced directly by the annihilation process ({\em i.e.} prompt photons or gluons produced at the partonic level)  and those that will emerge from the process of fragmentation into hadrons from both the final state quark-antiquark and the gluon from bremsstrahlung. This requires to resort to Monte-Carlo simulation tools, like {\sc Pythia}~\cite{Sjostrand:2014zea}.  The way we handle this  is discussed in the next section. Here we focus on the differential cross section at the partonic level and on the possible simplifications one may use to get approximate results. Eq.~\eqref{eq:dSVdEhard} is in principle all we need to determine the spectrum of prompt gluons or gammas. Its expression is however not very convenient, as it involves all the terms given in the Appendix~\ref{app:sv}. In practice, we may rely on rather simple approximations. The first is to replace the full expression of Eq.~\eqref{eq:dSVdEhard} by
\begin{equation}
\label{eq:diff_approx1}
\frac{d\sigma}{d\omega}\approx \left.\frac{d\sigma_{\rm FSR}}{d\omega}\right |_{m_q\neq 0}+\frac{d\sigma_{\rm VIB}^{(0)}}{d\omega}\;.
\end{equation}
In this expression, the first term is the differential cross section for emission of a gamma or gluon using the effective theory, that is the amplitudes  in Figs.~\ref{fig:3B-EFTa} and \ref{fig:3B-EFTb}. Basically, this contribution is equivalent to the process of Higgs decay studied in e.g. \cite{Drees:1990dq}. The second term is the differential cross section for VIB calculated from the amplitude of Fig.~\ref{fig:VIB}, where the {superscript in $\sigma_{\rm VIB}^{(
0)}$ refers to the cross section in the limit of a massless quark}, see Eq.~\eqref{eq:dVIB0} in Appendix \ref{app:limits}. The rationale is that VIB is mostly relevant in the limit $m_q \ll M_S$ (and provided the mediator is not much heavier than the DM particle). Less obvious is that this expression works pretty well for intermediate regimes. This is illustrated in Fig.~\ref{fig:InterpolationDiffxSR} where, for the specific choice of $M_S = 2$ TeV, $\MP  =2.4$ TeV and $m_q = m_t$,  we show the differential cross section for prompt gamma-ray emission based on the full calculation (solid black line) compared to the one obtained from the expression in Eq.~\eqref{eq:diff_approx1} (dotted blue line).  The two curves are clearly very close to each other. We should emphasize that to get a better matching we have shifted the energy $\omega$ in the differential cross section for the VIB contribution to $\omega + m_q^2/M_S$, so as to take into account the finite quark mass effect on the end-point of the gamma-ray spectrum. Modulo this, the correspondence between the two expressions is surprisingly good. 
This simple decomposition into FSR plus VIB suggests a further approximation, which is to express FSR in terms of the LO 2-body annihilation cross section time a factor that takes into account the emission of a gamma-ray or gluon by the final state quarks, 
\begin{equation}
\label{eq:split}
\frac{d\sigma_{\rm FSR}}{d\omega} \approx \sigma v_{q\overline{q}}\left(\frac{\alpha Q^2}{\pi}\right)\left\{ \mathcal{F}\left(\omega\right)\log\left(\frac{4M_S\left(M_S-\omega\right)}{m_q^2}\right) - {2 M_S\over \omega}\right\}\;.
\end{equation}
In this expression, the factor $\mathcal{F}$
is the standard splitting function for emission of gamma by a final state fermion,
\begin{align}
\mathcal{F}\left(\omega\right)=\frac{M_S^2+\left(M_S^2-\omega\right)^2}{M_S}\frac{1}{\omega}\;.
\end{align}
 In case of emission of a gluon, one must of course replace the factor $\alpha Q^2$ by $\alpha_s C_F$ in Eq.~\eqref{eq:split}. The splitting function captures the collinear divergences that arise in the limit $m_q \ll \omega \sim M_S$, hence for hard emission \cite{Birkedal:2004xn}.   Integrating over $\omega$ down to a cut-off energy $\omega_0$ leads to the characteristic Sudakov double logarithmic divergence $\propto \log(m_q^2/M_S^2) \log(m_q^2/\omega_0^2)$  which one can read in the expression of Eq.~\eqref{eq:sv_hard}. We have checked explicitly that the expression of   ${d\sigma_{\rm FSR}}/{d\omega}$ reduces to the expression of Eq.~\eqref{eq:split} in the limit $m_q \ll \omega \sim M_S$. {Doing so, we have also obtained, on top of the universal Sudakov term, the last term in Eq.~\eqref{eq:split}. This non-universal term is {\em a priori} subdominant  in the limit $m_q \ll \omega \sim M_S$, but keeping it gives a good matching to the exact result over broader range of energies.} This is shown in Fig.~\ref{fig:InterpolationDiffxSR} where the dashed red line correspond to the differential cross section obtained by summing the contribution from Eq.~\eqref{eq:split} to VIB (in the massless limit). The matching is not as good as in the previous approximation, Eq.~\eqref{eq:diff_approx1}, but for practical applications, it is much simpler and also more physically transparent. It also paves the way to the determination of the final gamma-ray spectrum from hadronisation.

\subsection{Gamma-ray spectra}

Our final goal is to obtain the gamma-ray spectrum from annihilation of a $S$ dark matter candidate, taking into account FSR and VIB emission both of gluons and gammas. One obvious way to proceed is to implement directly our differential 3-body process into {\sc Pythia}, by first building a Monte-Carlo distribution using {\em e.g.} {\sc
  CalcHEP}~\cite{Belyaev:2012qa} which then can be hadronised using {\sc Pythia}. This is the strategy we have used in the past for the case of coupling of the vector-like portal to light quarks \cite{Giacchino:2015hvk}, in which case we could altogether neglect the quark mass and so the 2-body annihilation process. 
Clearly this strategy also applies to heavy quarks in the limit $m_q \ll M_S \rightarrow 0$, provided the mediator is not much heavier than the DM candidate, see Fig.\ref{fig:InterpolationxSL}. In general, however, the fermion mass and the associated FSR may not be neglected.  The problem is 
that infrared and collinear singularities associated to FSR lead to sharp peaks in the Monte-Carlo distribution, which, for numerical convergence purpose, require to introduce a cut-off on the energy and, {\em a priori} on the emission angle of emitted gluon or gamma to obtain a reliable numerical output. Doing so for each DM candidate is cumbersome and CPU time consuming. Now, as  discussed at length in the previous sections, the total cross section(s) can be  decomposed into soft and hard gluon or gamma emission. So, the next idea is to implement separately soft and hard emissions. The latter is free of infrared and collinear divergences and so can be reliably calculated using first {\sc CalcHEP} to simulate the 3-body process and then {\sc Pythia} for hadronisation. 
One still has to deal with divergences from soft emission. This however, is a radiative correction to a 2-body process that can be handled directly by {\sc Pythia}, which has built-in simulation of FSR emission, including splitting of gluons, through Sudakov factors \cite{Sjostrand:2014zea}. 

\begin{figure}[h]
\centering
\includegraphics[width=8cm]{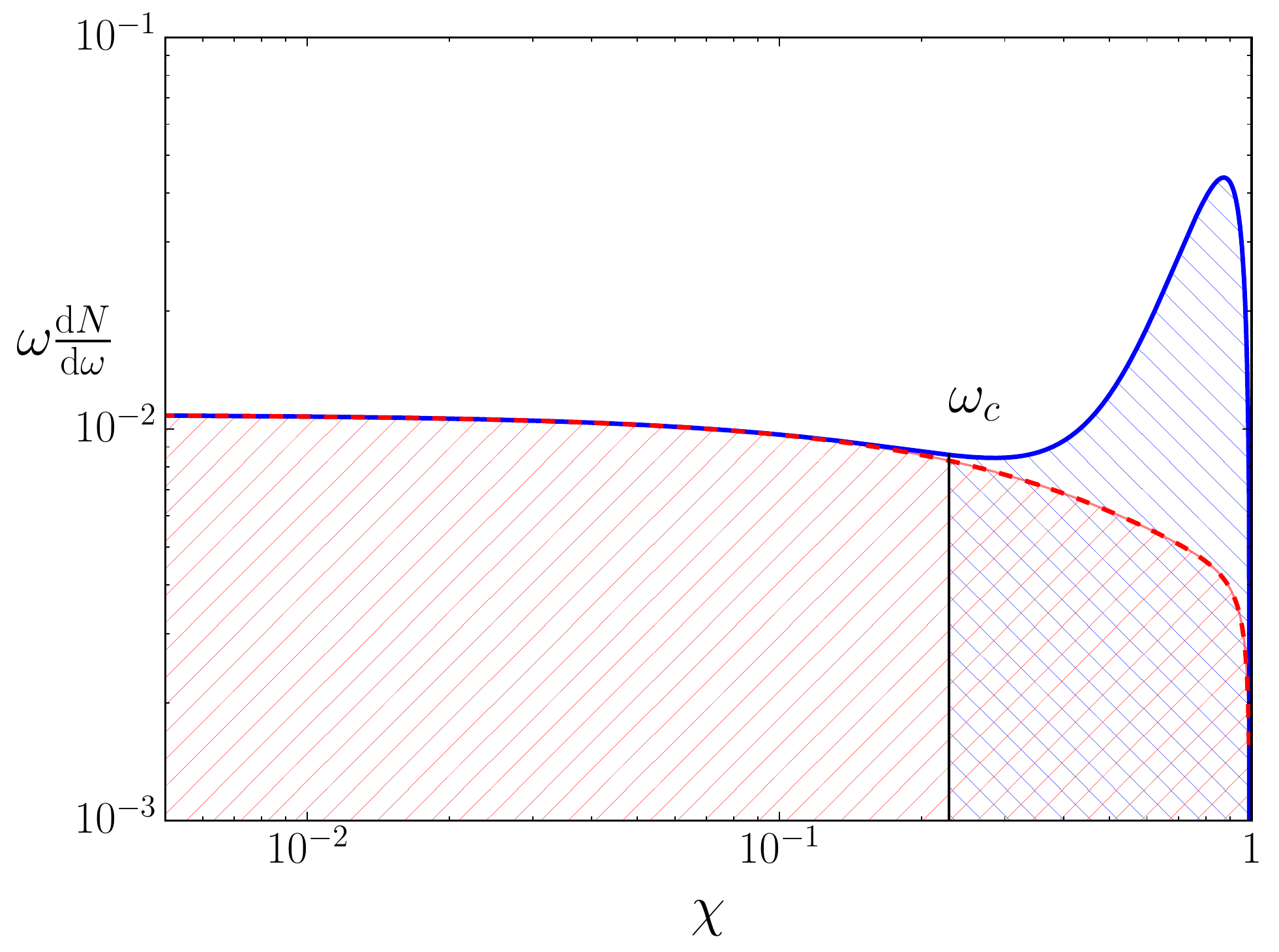}
\label{fig:omega0choiceL}
\caption{\label{fig:omega0choice} Left panel: Differential cross section of gluons in the full theory and the $r\rightarrow \infty$ one. The hatched red and blue are respectively the soft and hard contributions. Note that the crossed region is counted twice but the additional contribution is subtracted, see text.}  
\end{figure}

The separation between soft and hard modes however amounts to consider non-inclusive cross sections, which typically have Sudakov double logarithm divergences.  So a question is where to put the cut-off between soft and hard modes. Indeed, Eq.~\eqref{eq:soft_eff}  has a  Sudakov double logarithm $\propto \log(m_q^2/M_S^2) \log(m_q^2/\omega_0^2)$ typical of non-inclusive cross sections and which can be traced to the behavior of the differential cross section in the regime of collinear emission, so we should avoid taking the cut-off at low energies where the log diverges. Instead we  take it such that the FSR and VIB match smoothly. This is depicted in Fig.~\ref{fig:omega0choice} where for concreteness we show the spectrum of gamma-rays (at the partonic level) for a candidate with a strong VIB feature. The blue, solid line, is the spectrum obtained using the full theory, and the red, dashed line the one from final state radiation using the 2-body effective theory. The cut-off $\omega_c$ that separates high and low energy emission may be chosen as the energy where the differential cross section in the effective theory departs from the one in the full theory. 

Once $\omega_c$ is determined, we can generate  hard events
(with {\sc CalcHEP} and then {\sc Pythia} for hadronisation) 
using the full theory on one hand, and soft events coming from $q\bar q$ with FSR events (using {\sc Pythia} only) on the other hand. Using the theoretical expression for the differential cross section in the full theory, it is easy to generate (using {\sc CalcHEP})  gluons or gammas with an energy larger than the cut-off $\omega_c$. However, within {\sc Pythia8} we did not find a simple way to extract partonic level events with  gluons or photons with energy $\omega < \omega_c$. Instead, we have first generated with {\sc Pythia8} the complete gamma-ray spectrum from the 2-body process, including FSR. Next, we have simulated with {\sc CalcHEP} the distribution of hard gluons and gammas events with $\omega > \omega_c$ using the analytical cross section for the 2-body process with FSR. After hadronisation of this part with {\sc Pythia8}, we have subtracted the resulting gamma-ray spectrum  from the one obtained in the first step, to get only the soft part of the gamma-ray spectrum. Finally we have added back the hard part from the full theory, which includes VIB effects, to get the final gamma-ray spectrum. As $S$ annihilation proceeds through several final states ($q\overline{q}+\gamma+g$, $\gamma\gamma$, $gg$), the total photon spectrum, ${dN_\gamma^{\text{tot}}}/{dE_\gamma} $, is finally given by the sum of the photon spectrum originating from each final state (${dN_\gamma^{t\overline{t}+\gamma+g}}/{dE_\gamma}$, ${dN_\gamma^{\gamma\gamma}}/{dE_\gamma}$ and ${dN_\gamma^{gg}}/{dE_\gamma}$  respectively), weighted with their respective branching ratios. To distinguish prompt emission gluons and gammas of energy $\omega$, we use $E_\gamma$ for the energy of the gamma-rays in the final spectrum. The normalization is with respect to the total full inclusive annihilation cross section. 
\begin{figure}[th]
\centering
\begin{subfigure}[b]{0.45\textwidth}
\includegraphics[width=\textwidth]{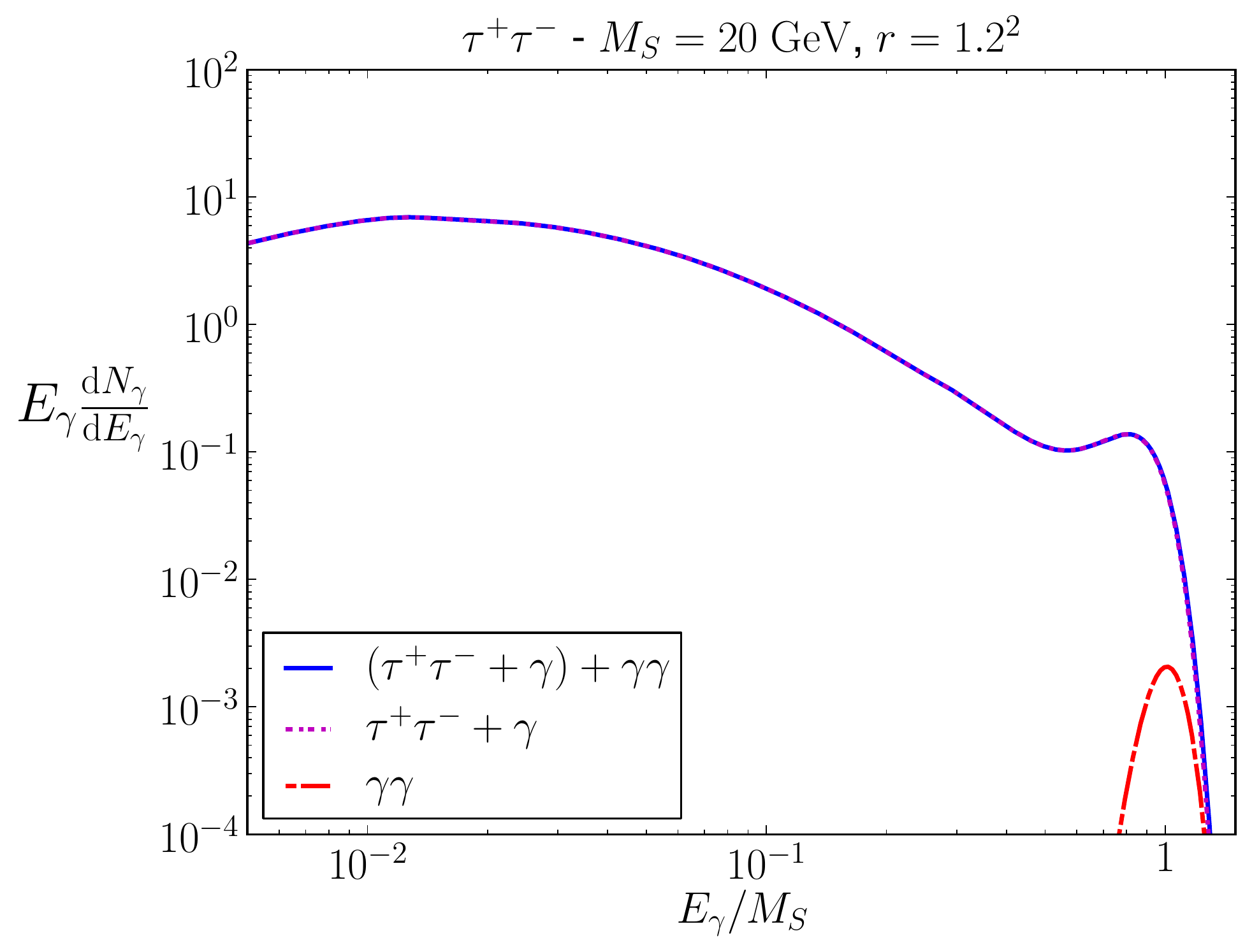}
\end{subfigure}
\quad \quad
\begin{subfigure}[b]{0.45\textwidth}
\includegraphics[width=\textwidth]{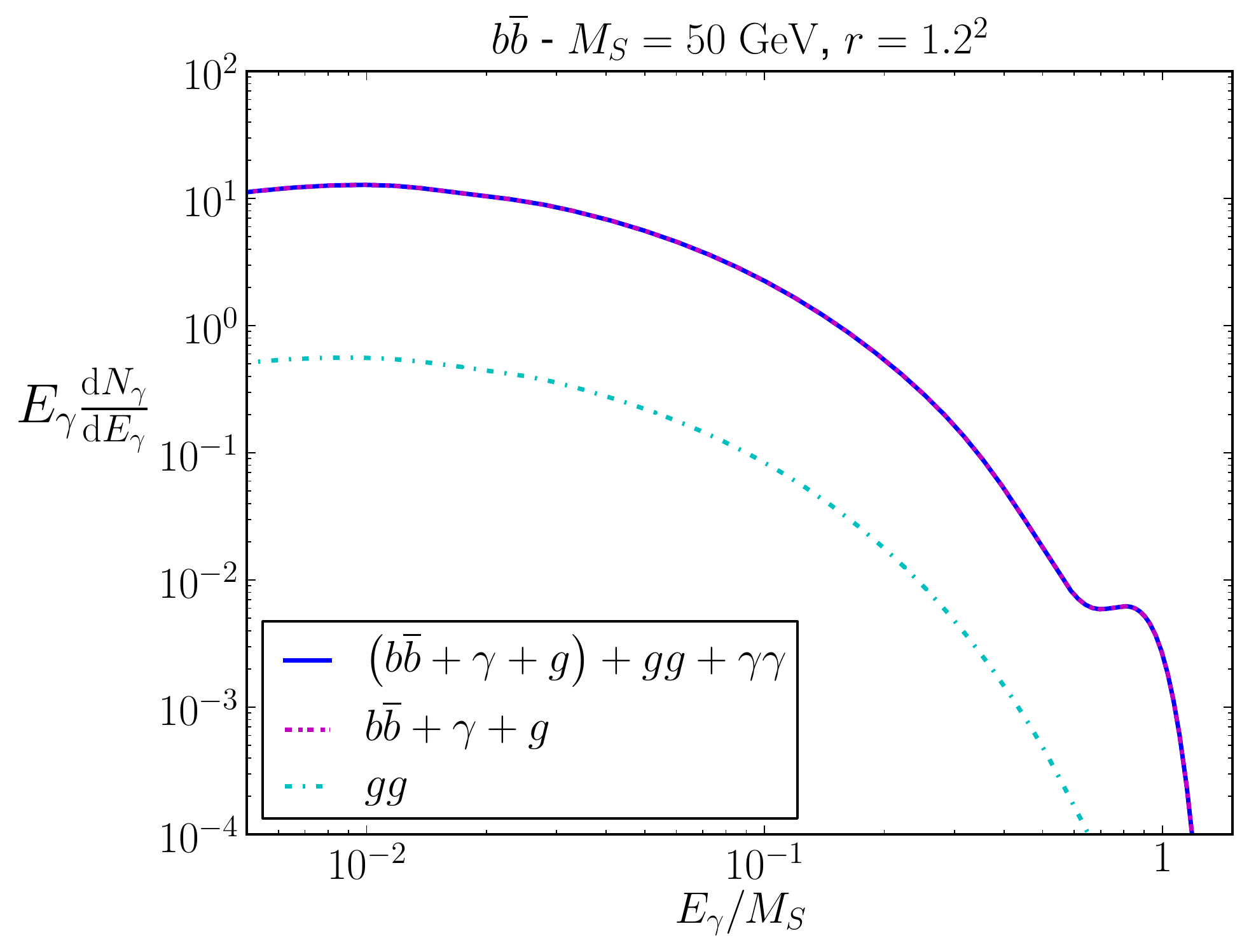}
\end{subfigure}
\caption{\label{fig:bandtau} Gamma ray spectrum from annihilation in {a $\tau^+\tau^-$ (left) or $b\bar b$ (right) models.  In the left-panel, the dotted (purple) line is the total spectrum from $\gamma$-ray bremsstrahlung together with $\tau^+\tau^-$ hadronization and the dot-dashed (red) curve is the 1-loop gamma-ray line. We have assumed that the resolution on the gamma ray energy is gaussian distributed with a relative error of $10 \%$. The solid (blue) line is the total spectrum. The right panel is for annihilation into $b\bar b$. In this case, bremsstrahlung includes radiation of a gluon and the dot-dashed (light blue) line is from the hadronization of the 1-loop gluon line. These show that the 1-loop monochromatic emission is negligible and that the feature at high gamma-ray energies is entirely due to the virtual internal bremsstrahlung contributions.} }  
\end{figure}
\begin{figure}[th!]
\centering
\begin{subfigure}[b]{0.45\textwidth}
\includegraphics[width=\textwidth]{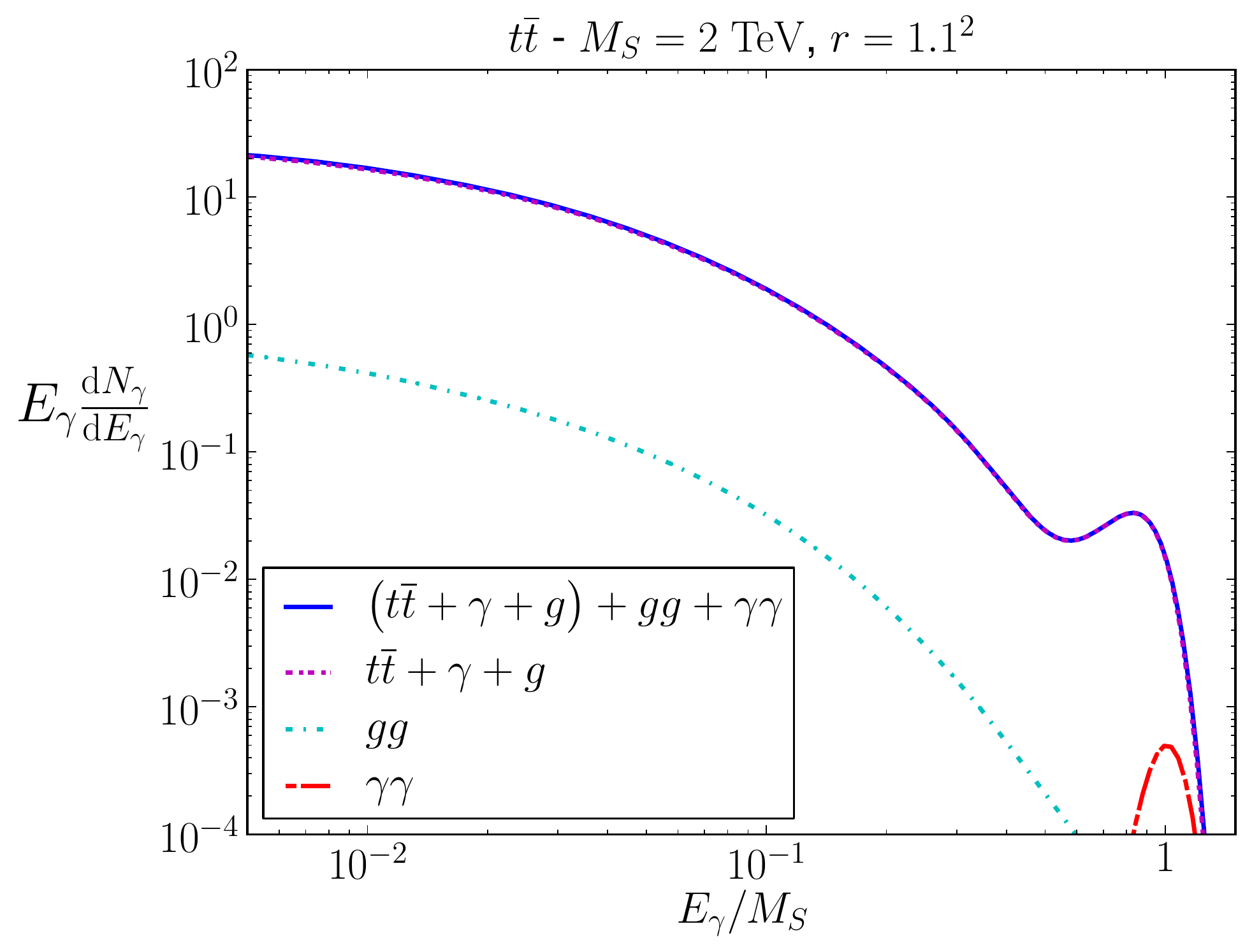}
\end{subfigure}
\quad \quad
\begin{subfigure}[b]{0.45\textwidth}
\includegraphics[width=\textwidth]{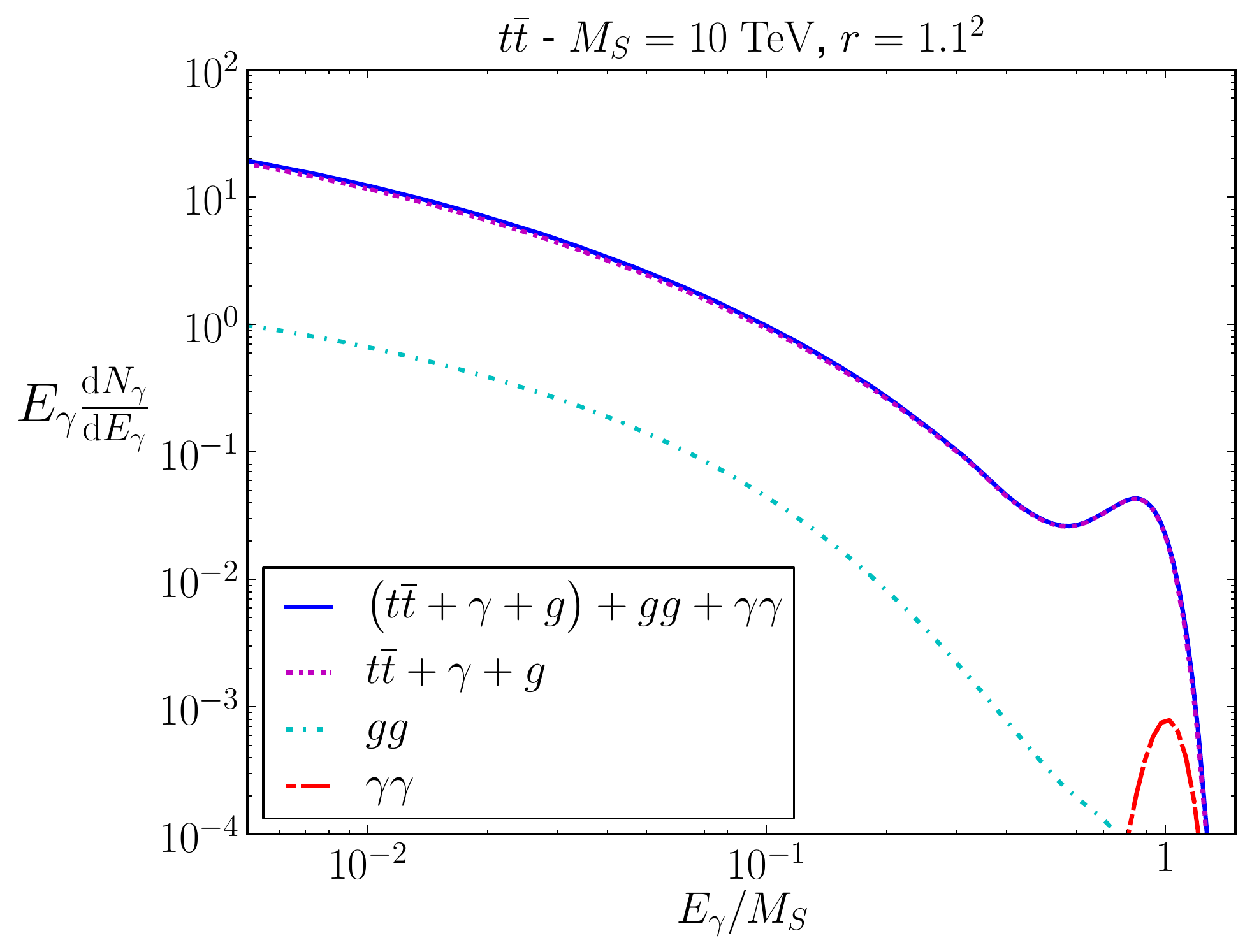}
\end{subfigure}
\caption{\label{fig:tt}Characteristic gamma-ray spectrum from annihilation into $t\bar t$. {The meaning of the various curves are the same as in Fig.~\ref{fig:bandtau}.}}
\end{figure}

We may now consider, for the sake of illustration, different SM fermionic final states for which the fermion mass may be a relevant parameter. Concretely, we consider the $\tau^+\tau^-$ as well as the $b\bar b$ (Fig.~\ref{fig:bandtau}) and $t\bar t$ (Fig.~\ref{fig:tt}) channels. We show spectra for  fixed $(m_q/M_S)^2 \equiv z$ ratios, $z=0.1^2$, $r=1.2^2$, adjusting both the dark matter mass $M_S$ and mediator mass $\MP$. Each spectrum is generated following the procedure depicted above, with a specific cut-off separating emission of soft and hard gluons or gammas. We have included for completeness the contribution from annihilation at one-loop into two gluons and into $\gamma\gamma$,  assuming a gamma-ray detector with resolution $\Delta E_\gamma/E_\gamma = 10 \%$. The parameters of the DM model are chosen so as to illustrate the possible presence of a feature in the final gamma-ray spectrum, not taking into account other possible constraints (relic abundance, direct, indirect and collider constraints). In that respect, most relevant are the two plots of Fig.\ref{fig:tt} with coupling of DM to the top quark, which corresponds to actual DM candidates. The phenomenology of such DM candidates are discussed in details in \cite{Colucci:2018vxz}. 
\section{Summary and conclusions}
\label{sec:con}

We have studied further radiative corrections to a simple DM scenario, in the form of a real scalar particle annihilating into SM fermions through a heavy vector-like fermion. This topic has been already covered in several phenomenological studies \cite{Toma:2013bka,Giacchino:2013bta,Giacchino:2014moa,Ibarra:2014qma,Giacchino:2015hvk} that focused on coupling to light quarks or leptons. Of particular relevance in this regime is the helicity suppression of the 2-body annihilation cross section, akin to p-wave suppression of Majorana dark matter annihilating into SM chiral fermions \cite{Goldberg:1983nd,Bergstrom:1989jr,Flores:1989ru}. This implies in particular that radiative corrections, in the form of one-loop annihilation into two photons (or gluons) and so-called virtual internal bremsstrahlung may play a significant role both in determining the relic abundance and for indirect searches. Due to infrared and collinear divergences that affect bremsstrahlung of massless gauge bosons, the extension of these results to heavy quarks (or annihilation into a $\tau^+\tau^-$ pair) poses some technical problems, which we have tried to overcome in the present work. The motivation was manifold, but the main aim was to try and test simple approximations that could be then applied for phenomenological studies, in particular to the case of top-philic coupling, a topic of much interest, in particular from the perspective of simplified DM searches at the LHC. Such study has been the object of a separate publication \cite{Colucci:2018vxz} (see also \cite{Baek:2016lnv,Baek:2017ykw,Keung:2017kot,Arina:2018zcq}). 
Here, we focused on technical aspects of the calculations. In particular, following a proposal of \cite{Bringmann:2015cpa}, we have adapted an effective approach suited for emission of soft gamma or gluons and that circumvent several unnecessary steps in the regularization of infrared divergences. From an effective approach perspective, much of the calculations map to the equivalent problem of radiative corrections to Higgs decay into SM fermions \cite{Drees:1990dq,Braaten:1980yq}.  This approach, which is quite systematic, albeit pedestrian, has, we believe, also a pedagogical value. In particular, it illustrates in a simple framework how the cancellation of infrared and collinear divergence takes place in the calculation of a total cross section, in agreement with standard quantum field theory theorems. At the end of the day, our main results include a full, explicit but unpractical expression for the total annihilation cross section for $S$ DM into SM fermions at NLO in $\alpha_s$ (or $\alpha$). 
Building on this full calculation, we have studied several simple approximations, both to the total and differential cross sections. The take home lesson is that the simple approximations discussed in the bulk of this article are well suited for phenomenological studies, as discussed in details for the case of coupling to the top quark \cite{Colucci:2018vxz}. 
{As an invitation to \cite{Colucci:2018vxz}, we briefly discuss here the impact of radiative corrections on the relic abundance for such top-philic scenario. In Figs.~\ref{fig:TwoApproxFig}, the solid (red) lines border, in the $\left(M_S,{M_\psi}/{M_S}-1\right)$ plane, delimitate the DM candidates whose abundance match the cosmological observations. As the model has only three parameters, to each candidate between the two solid (red) line corresponds a viable candidate and thus a specific value of its Yukawa coupling. In the left panel, Fig.~\ref{fig:TLApprox}, we report the absolute difference between the tree-level or leading order annihilation cross section $\sigma v_{q\overline{q}}$ and the full expression we have obtained taking into account radiative corrections (NLO) or $1-\sigma v_{q\overline{q}}/\sigma v_{\text{NLO}}$. Clearly, radiative corrections begin to be substantial for $m_q/M_S = m_\text{top}/ 1\, \mathrm{TeV}\approx 0.2 $. One also  sees that radiative corrections are substantial for large values of $M_S$ and relatively degenerate values of $M_\psi$. This reflects the fact that the radiative processes become more and more dominated by VIB emission in this regime, as already emphasized and explained at several places in the literature, both for Majorana \cite{Bringmann:2007nk,Bringmann:2012vr,Bell:2011eu,Garny:2013ama,Garny:2015wea} and scalar DM \cite{Toma:2013bka,Giacchino:2013bta,Giacchino:2014moa,Ibarra:2014qma}. We refer the reader to Section \ref{sec:discussion} for more details, were one can also find the origin of specific features. For instance the narrow funnel region seen in Fig.~\ref{fig:TLApprox} for large $M_S$ corresponds to parameters for which the NLO corrections change sign. One sees from Fig.~\ref{fig:InterpolationxSL} that QCD correction are negative in the regime for which the 2-body cross section is dominant, leading to a decrease of the total annihilation cross section, while as $M_\psi/M_S$ increases, VIB becomes eventually dominant over
$\sigma v_{q\bar q}$.
In Fig.~\ref{fig:WorstApprox}, we show the improvement we obtain taking into account only VIB in the massless limit, namely $1-\left(\sigma v_{q\overline{q}}+ \sigma v_{\text{VIB}}^{(0)}\right)/\sigma v_{\text{NLO}} $. Using instead Eq.~(\ref{eq:InterpolationBetter}) would have given an error that is less than $10\%$ over the whole $\left(M_S,{M_\psi}/{M_S}-1\right)$ plane. This illustrates, for the specific problem of determining the abundance, both the relevance of radiative corrections and the benefit of using approximate expressions, key results of the present work.\footnote{{The residual errors at low mass $M_S<300\, \mathrm{GeV}$ are due to bound state formation effects. In that region, the simplest cure is to use the tree-level cross section, see the discussion in Section \ref{sec:discussion}.}} Further aspects, like the impact of bremsstrahlung for indirect detection, are discussed extensively in Ref.\cite{Colucci:2018vxz}, to which we refer for further details.
\begin{figure}[t]
\centering
\begin{subfigure}[b]{8cm}
\includegraphics[width=\textwidth]{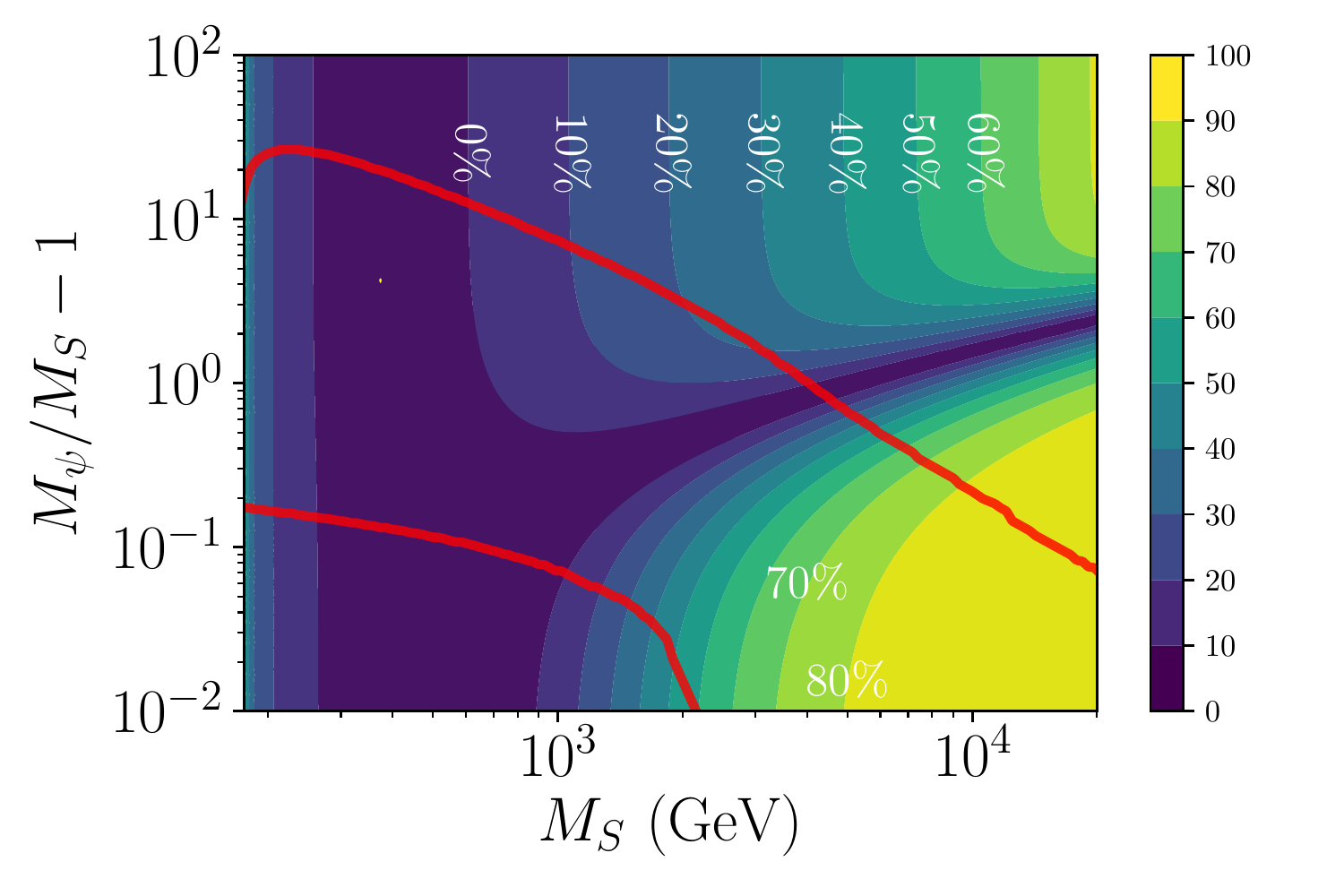}
\subcaption{
\label{fig:TLApprox}}
\end{subfigure}
\quad \quad
\begin{subfigure}[b]{8cm}
\includegraphics[width=\textwidth]{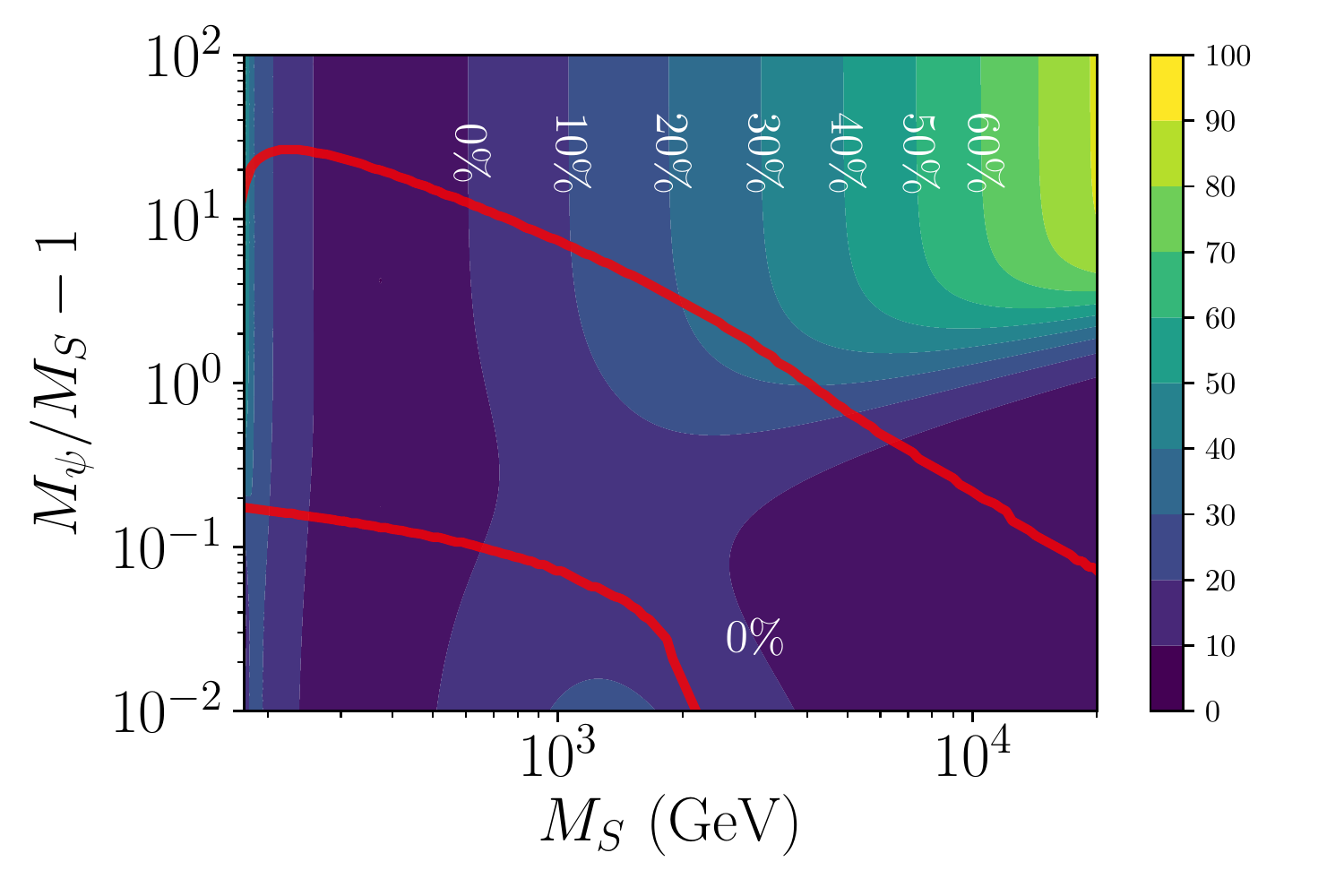}
\subcaption{\label{fig:WorstApprox}}
\end{subfigure}
\caption{{Both panels: the DM candidates within the solid (red) lines in the plane $\left(M_S,{M_\psi}/{M_S}-1\right)$ have an abundance that matches the cosmological observations \cite{Colucci:2018vxz}. The contours in the left panel give the relative error between the tree-level annihilation cross section $\sigma v_{q \overline{q}}$  and its full expression at NLO, $\left|\sigma v_{\text{NLO}}-\sigma v_{q\overline{q}}\right|/\sigma v_{\text{NLO}}$. The right panel gives instead $\left|\sigma v_{\text{NLO}}-\sigma v_{q\overline{q}}- \sigma v_{\text{VIB}}^{(0)}\right|/\sigma v_{\text{NLO}} $. See text.}}
\label{fig:TwoApproxFig}
\end{figure}

\appendix
\section{$S_0(\chi)$ function} \label{app:sv}

 The function $S_0(\chi)$ consists of a combination of several polynomials $P_i\left(\chi \right)$:
\begin{eqnarray}
 \label{eq:S0func}
S_0(\chi)&=&\frac{-1}{2\Delta^2}\Bigg[\frac{P_1\left(\chi\right)}{\beta\left(\Delta-\chi\right)^2}+\frac{P_2\left(\chi\right)}{\Delta\beta\left(\Delta-\chi\right)}+\frac{\beta}{1-\beta^2}\left(\frac{1}{\Delta^3}\left( \beta^2 P_3\left(\chi\right)+P_4\left(\chi\right)\right)+P_5\left(\chi\right)\right) \nonumber\\
&+&\frac{1}{\Delta^2-2\Delta+\left(1-\beta^2\right)\chi^2}\left(\frac{\beta}{\Delta}P_6\left(\chi\right)+\frac{1}{\Delta\beta} P_7\left(\chi\right)+\frac{1}{\Delta^3}\frac{\beta}{1-\beta^2}\bigg(\beta^4 P_8\left(\chi\right)+P_9\left(\chi\right)+\beta^2 P_{10}\left(\chi\right)\bigg)\right) \nonumber\\ 
&+&\log\frac{1+\beta}{1-\beta}\left(\frac{P_{11}\left(\chi\right)}{\Delta}+\frac{P_{12}\left(\chi\right)}{\Delta-2\chi}+P_{13}\left(\chi\right)\right)+\log\frac{\Delta-\left(1-\beta\right)\chi}{\Delta-\left(1+\beta\right)\chi} \bigg(\frac{\Delta^2}{\left(\Delta-\chi\right)^3}P_{14}\left(\chi\right) \nonumber\\ 
&+&\frac{P_{15}\left(\chi\right)}{\Delta}+\frac{P_{16}\left(\chi\right)}{\Delta^2}\frac{1}{\Delta-2\chi}+\frac{P_{17}\left(\chi\right)}{\Delta^2}\frac{1}{\Delta-\chi}+\frac{P_{18}\left(\chi\right)}{\Delta}\frac{1}{\left(\Delta-\chi\right)^2}\bigg)\Bigg]\;,
\end{eqnarray}
where we have used the shorthand $\Delta \equiv 1+r-z$ for the factor originated from the propagator of the mediator field $\psi$. The $P_i$ themselves are polynomials in $\chi$: 
\begin{eqnarray} 
\nonumber
P_1\left(\chi\right)&=&2\left(r^3+r\left(1-z\right)^2+2\left(1-z\right)^3+2r^2z\right)\chi-4\left(r^2+\left(1-z\right)^2\right)\chi^2\nonumber\\ 
P_2\left(\chi\right)&=&2\Delta\left(5-4r+3r^2-6z+z^2\right)\chi-4\left(3-r+4r^2-4z+z^2\right)\chi^2+8r\chi^3\nonumber\\ 
P_3\left(\chi\right)&=&6\Delta^2z\left(2r-\Delta\right)\chi+4\Delta z\left(2r-\Delta\right)\chi^2-16z\left(2r-\Delta\right)\chi^3\nonumber\\ 
P_4\left(\chi\right)&=&4\left(r^2-2r\left(2-z\right)-5\left(1-z\right)^2\right)\Delta^2z-2\left(3r-11\left(1-z\right)\right)\Delta^2z\chi\nonumber\\ 
&&-20\Delta z\left(2r-\Delta\right)\chi^2+16z\left(2r-\Delta\right)\chi^3 \nonumber\\
P_5\left(\chi\right)&=&16\left(1-z\right)z \nonumber\\
P_6\left(\chi\right)&=&-2\Delta\left(1-z\right)\left(\left(1+r\right)^2+2rz-z^2\right)\chi+2\left(r\left(2+r\right)+\left(1-z\right)^2\right)\Delta \chi^2 \nonumber \\
&&+4\left(3-3r+2r^2-2\left(2-r\right)z+z^2\right)\chi^3-8r\chi^4 \nonumber\\
P_7\left(\chi\right)&=& -2\Delta\left(4r^3+2r\left(1-z\right)^2-r^2\left(1+z\right)+\left(1-z\right)^2\left(7-3z\right)\right)\chi+2\Delta\big(r\left(-6+13r\right)\nonumber\\
&&+\left(1-z\right)\left(13-5z\right)\big)\chi^2-4\left(3r+6r^2-2\left(2+r\right)z+z^2\right)\chi^3+8r\chi^4 \nonumber\\
P_8\left(\chi\right)&=&6\Delta^2z\left(2r-\Delta\right)\chi^3+4\Delta z\left(2r-\Delta\right)\chi^4-16z\left(2r-\Delta\right)\chi^5  \nonumber\\
P_9\left(\chi\right)&=& 4\Delta^4z\left(-r^2+\left(1-z\right)^2+2rz\right)+4\Delta^3z\left(3r^2-7\left(1-z\right)^2-r\left(4+z\right)\right)\chi \nonumber \\
&&+4\left(r^2+r\left(8-6z\right)+7\left(1-z\right)^2\right)\Delta^2z\chi^2-2\left(25r-17\left(1-z\right)\right)\Delta^2 z\chi^3+52\Delta z\left(2r-\Delta\right)\chi^4\nonumber \\
&&-16z\left(2r-\Delta\right)\chi^5 \nonumber\\
P_{10}\left(\chi\right)&=&-4\Delta^3z\left(r^2-\left(1-z\right)^2-rz\right)\chi+4\Delta^2z\left(3r^2-3\left(1-z\right)^2-2rz\right)\chi^2\nonumber\\
&&+4\Delta^2\left(1+3r-z\right)z\chi^3-56\Delta z \left(2r-\Delta\right)\chi^4+32z\left(2r-\Delta\right)\chi^5\nonumber
\end{eqnarray}
\begin{eqnarray}
P_{11}\left(\chi\right)&=&3-r\left(1+\left(3-r\right)r\right)-z+r\left(8+r\right)z-\left(5+9r\right)z^2+3z^3-2\Delta\left(2r-\Delta \right)\chi \nonumber\\
P_{12}\left(\chi\right)&=&-19-r\left(15-\left(3-r\right)r\right)+29z+\left(4-r\right)rz-\left(7-9r\right)z^2-3z^3\nonumber\\
&+&\left(4\left(13+r\left(2+r\right)\right)-2\left(19+r\right)z-6z^2\right)\chi-4\left(7+r-z\right)\chi^2\nonumber\\
P_{13}\left(\chi\right)&=&4\left(4-z\left(2+z\right)\right)-4\left(4-z\right)\chi\nonumber\\
P_{14}\left(\chi\right)&=&-r^3-r\left(1-z\right)^2-2\left(1-z\right)^3-2r^2z+2\left(r^2+\left(1-z\right)^2\right)\chi\nonumber\\
P_{15}\left(\chi\right)&=&3-r\left(1+\left(3-r\right)r\right)-z+r\left(8+r\right)z-\left(5+9r\right)z^2+3z^3-2\Delta\left(2r-\Delta\right)\chi\nonumber\\
P_{16}\left(\chi\right)&=&-\Delta^2\left(7r^2-r^2\left(5+17z\right)+r\left(-15+\left(34-21z\right)z\right)-\left(1-z\right)\left(3-z\left(8-11z\right)\right)\right) \nonumber\\
&+& 2\Delta\left(2\left(1+r\right)\left(4-3r\right)\left(1-3r\right)-\left(3-5\left(12-5r\right)r\right)z-2\left(11+21r\right)z^2+17z^3\right)\chi \nonumber\\
&-&4\left(21r^3+r^2\left(7-25z\right)+\left(1-z\right)^2\left(13+9z\right)-r\left(1-z\right)\left(1-13z\right)\right)\chi^2 \nonumber\\
&+&16\left(1+6r^2+r\left(3-4z\right)-z^2\right)\chi^3-32r\chi^4\nonumber\\
P_{17}\left(\chi\right)&=&
\Delta^2\left(4r^3-13r^2z-\left(1-z\right)\left(4-3\left(1-z\right) z\right)-2r\left(4-z\left(5-2z\right)\right)\right)\nonumber\\
&-&2\Delta\left(2\left(1-r\right)\left(1+r\right)\left(2-5r\right)-\left(2-3\left(9-5r\right)r\right)z-\left(10+19r\right)z^2+8z^3\right)\chi \nonumber\\
&+&4\left(11r^3+r^2\left(5-14z\right)+\left(1-z\right)^2\left(7+4z\right)+r\left(1+\left(4-5z\right)z\right)\right)\chi^2\nonumber\\
&-&8\left(1+6r^2+r\left(3-4z\right)-z^2\right)\chi^3+16r\chi^4 \nonumber\\
P_{18}\left(\chi\right)&=&\Delta^3\left(1-r\left(4-7r\right)-\frac{4r^2\left(1+r\right)}{\Delta}+2z+4rz-3z^2\right)+2\Delta\big(-5\Delta^2\left(1+z\right) \nonumber\\
&+&r\left(13+8r-5r^2-13\left(1-r\right)z\right)\big)\chi+2\big(11r^3+r^2\left(5-13z\right)+\left(1-z\right)^2\left(7+5z\right) \nonumber\\
&+&r\left(1+\left(6-7z\right)z\right)\big)\chi^2+4\left(1+6r^2+r\left(3-4z\right)-z^2\right)\chi^3+9r\chi^4
\end{eqnarray}

\section{Useful limiting behaviors} \label{app:limits}

In the limit $z  = m_q^2/M_S^2 \rightarrow 0$, the complex expressions of \ref{app:sv} reduce to 
\begin{eqnarray}
S_0(\chi)\vert_{z=0} &=&\frac{1-\chi}{\left(\Delta-2\chi\right)\left(\Delta-\chi\right)^3}\left(\frac{\left(\Delta-\chi\right)}{\Delta}\chi\left(\Delta^2-2\Delta \chi+2\chi^2\right)\right.\nonumber\\
&& \left.+ \frac{\Delta}{2}\left(\Delta-2\chi\right)^2\log\frac{\Delta-2\chi}{\Delta}\right)\;.
\end{eqnarray}
For annihilation in an s-wave, the LO cross section vanishes $\sigma v_{q \bar q}\vert_{z=0} = 0$, so all the complications induced by IR divergences drop, and one recovers the known expressions for VIB for massless fermions
\begin{eqnarray}
\label{eq:dVIB0}
{d\sigma^{(0)}_{\rm VIB} \over d \omega}
&=&\frac{N_c y_f^4}{8\pi M_S^3}\frac{\alpha_S C_F}{\pi}\frac{1-\chi}{1+r}\frac{1}{\left(1+r-2\chi\right)\left(1+r-\chi\right)^3}.\nonumber\\
&&\Big(2\left(1+r-\chi\right)\chi\left(\left(1+r\right)^2-2\left(1+r\right)\chi+2\chi^2\right)+\nonumber\\
&&-\left(1+r\right)^2\left(1+r-2\chi\right)^2\log\frac{1+r}{1+r-2\chi}\Big)\;,
\end{eqnarray}
and
\begin{eqnarray}
\label{eq:VIB0}
\sigma v_{\text{VIB}}^{(0)}&=&\frac{N_c y_f^4}{8\pi M_S^2}\frac{\alpha_S C_F}{\pi}\left\{\left(r+1\right)\left(\frac{\pi^2}{6}-\log^2\frac{1+r}{2r}-2\text{Li}_2\left(\frac{1+r}{2r}\right)\right)\right.\nonumber\\
&&\left.+\frac{4r+3}{r+1}+\frac{4r^2-3r-1}{2r}\log\frac{r-1}{r+1}\right\}\;.
\end{eqnarray}
In the limit $r\rightarrow 1$,
\begin{equation}
\label{eq:zeroVIB}
\sigma v_{\text{VIB}}^{(0)} \rightarrow \frac{N_c y_f^4}{8\pi M_S^2}\frac{\alpha_S C_F}{\pi}\left({7\over 2} - {\pi^2\over 3}\right)\;.
\end{equation}

In the opposite limit, $r = \MP/M_S \gg 1$,   VIB may be neglected.  With a factor of $\sigma v_{q \bar q}$ instead of the tree level decay rate of the Higgs, we recover the  expression of \cite{Drees:1990dq} (see erratum of \cite{Drees:1990dq}) 
\begin{align}
\label{eq:drees_NLO}
\sigma v_{q\bar q}^{\text{NLO}}=\sigma v_{q\overline{q}}\frac{C_F\alpha_S}{\pi}\left\{\frac{A\left(\beta_0\right)}{\beta_0}+\frac{3+34\beta_0^2-13\beta_0^4}{16\beta_0^3}\log\frac{1+\beta_0}{1-\beta_0}+\frac{3}{8\beta_0^2}\left(-1+7\beta_0^2\right)\right\}\;,
\end{align}
where
\begin{align}
A\left(\beta_0\right)&=\left(1+\beta_0^2\right)\left[4\text{Li}_2\left(\frac{1-\beta_0}{1+\beta_0}\right)+2\text{Li}_2\left(-\frac{1-\beta_0}{1+\beta_0}\right)-3\log\frac{2}{1+\beta_0}\log\frac{1+\beta_0}{1-\beta_0}-2\log\beta_0\log\frac{1+\beta_0}{1-\beta_0}\right]\nonumber\\
&-3\beta_0\log\frac{4}{1-\beta_0^2}-4\beta_0\log\beta_0\;.
\end{align}

\section*{Acknowledgements}
The work of M.T. is supported by the F.R.S.-FNRS, the IAP P7/37, the IISN convention 4.4503.15 and the Excellence of Science (EoS) convention 30820817. The work of F.G. and J.V. is supported by the F.R.I.A./Fonds pour la formation \`a la Recherche dans l'Industrie et dans l'Agriculture (FNRS). S.C. would like to thank the Service de Physique Th\'eorique at ULB for hospitality and support.

\bibliographystyle{JHEP}
\bibliography{ScalarDM}

\end{document}